\documentclass[12pt]{iopart}

\usepackage{iopams,setstack,graphicx}
\usepackage{cite}
\usepackage{color}
\usepackage{undertilde}

\eqnobysec

\newcommand{\mean}[1]{\left < #1 \right >}

\newcommand{\abs}[1]{\left | #1 \right |}

\newcommand{\eqnlabel}[1]{\refstepcounter{equation}\label{#1}\addtocounter{equation}{-1}}
\newcommand{\eqnref}[1]{(\ref{#1})}

\newcounter{Aeqnval}
\def\Anumparts{\addtocounter{equation}{1}%
     \setcounter{Aeqnval}{\value{equation}}%
     \setcounter{equation}{0}%
     \def\theequation{\ifnumbysec
     \Alph{section}.\arabic{Aeqnval}{\it\alph{equation}}%
     \else\arabic{Aeqnval}{\it\alph{equation}}\fi}}

\def\endAnumparts{\def\theequation{\ifnumbysec
     \Alph{section}.\arabic{equation}\else
     \arabic{equation}\fi}%
     \setcounter{equation}{\value{Aeqnval}}}

\newcommand{\fourier}[2]{{\hat{#1}}_{#2}}
\newcommand{\nabcp}{\utilde{\nabla}}

\begin{document}


	\title[Self-Propelled Particles with Selective Attraction-Repulsion Interaction]{Self-Propelled Particles with Selective Attraction-Repulsion Interaction - From Microscopic Dynamics to Coarse-Grained Theories}
	\author{R Gro{\ss}mann$^1$, L Schimansky-Geier$^1$ and P Romanczuk$^{2}$}

	\address{$^1$ Department of Physics, Humboldt-Universit\"at zu Berlin, Newtonstr. 15, 12489 Berlin, Germany}
	\address{$^2$ Max Planck Institute for the Physics of Complex Systems, N\"othnitzerstr. 38, 01187 Dresden, Germany}

	\ead{prom@pks.mpg.de}

	\begin{abstract}
	In this work we derive and analyze coarse-grained descriptions of self-propelled particles with selective attraction-repulsion interaction, where individuals may respond differently to their neighbours depending on their relative state of motion (approach versus movement away).  Based on the formulation of a nonlinear Fokker-Planck equation, we derive a kinetic description of the system dynamics in terms of equations for the Fourier modes of the one-particle density function. This approach allows effective numerical investigation of the stability of solutions of the nonlinear Fokker-Planck equation. 

 Further on, we also derive a hydrodynamic theory by performing a closure at the level of the second Fourier mode of the one-particle density function. We show that the general form of equations is in agreement with the theory formulated by Toner and Tu. The stability of spatially homogeneous solutions is analyzed and the range of validity of the hydrodynamic equations is quantified. 

 Finally, we compare our analytical predictions on the stability of the homogeneous solutions with results of individual-based simulations. They show good agreement for sufficiently large densities and non-negligible short-ranged repulsion. The results of the kinetic theory for weak short-ranged repulsion reveal the existence of a previously unknown phase of the model consisting of dense, nematically aligned filaments, which cannot be accounted for by the present hydrodynamics theory of the Toner and Tu type for polar active matter. 
	\end{abstract}	
		
	

	\maketitle


	\section{Introduction}

In recent decades, there has been an increased research focus on far-from-equilibrium systems in biology and physics which is referred to as ``active matter''. 
The relevant length scales of such systems span several orders of magnitude. They range from the \mbox{(sub-)}micrometer scale governing the dynamics of individual active units in motility assays in-vitro \cite{schaller_polar_2010} and the actin cortex in-vivo \cite{salbreux_actin_2012}, via the mesoscopic length scales of interest in collective dynamics of large bacterial ensembles \cite{wensink_meso-scale_2012,peruani_collective_2012} and artificial self-propelled particles \cite{theurkauff_dynamic_2012}, 
up to the macroscopic scales of driven granular matter \cite{kudrolli_swarming_2008,deseigne_collective_2010} or flocks of birds \cite{ballerini_interaction_2008}, schools of fish \cite{lopez_behavioural_2012}, and swarms of insects \cite{buhl_disorder_2006,bazazi_nutritional_2011}, where the spatial dimensions can be in the order of kilometers. 

Despite the apparent variety, all these systems share the fundamental property of local uptake and/or conversion of internal energy into kinetic energy of motion by its individual units.  This -- together with additional interactions between those units -- distinguishes these systems from related equilibrium systems and yields fascinating examples of self-organization and collective dynamics. 

The question of universal properties of such active matter systems from the statistical physics point of view is a vibrant research field. To a large extent, it was initiated by the numerical study of a minimal, individual-based model of active matter published by Vicsek \etal in 1995 \cite{vicsek_novel_1995}. 
Shortly after this publication, Toner and Tu made a seminal contribution by formulating the hydrodynamic equations of polar active matter at largest relevant length and time scales purely based on symmetry arguments \cite{toner_long-range_1995,toner_flocks_1998}. The analysis of these generic equations, as well as their counterparts for nematic order, improved our understanding of the fundamental properties of active matter, such as the existence of long range order or giant number fluctuations \cite{toner_hydrodynamics_2005,ramaswamy_mechanics_2010}.

However, the direct derivation of a hydrodynamic theory of the Toner and Tu type from microscopic models of active matter was a long standing problem. Only recently, such a link between microscopic parameters determining the dynamics of individual active units and parameters governing the macroscopic flow of active matter was established by formulating kinetic equations for minimal models of self-propelled particles with velocity-alignment \cite{bertin_boltzmann_2006,bertin_microscopic_2009,ihle_kinetic_2011,peshkov_continuous_2012} and self-propelled aligning rods \cite{peshkov_nonlinear_2012}. Furthermore, coarse-grained descriptions for active particles with variable speeds and velocity-alignment were derived in \cite{farell_marchetti_2012,grossmann_active_2012,romanczuk_collective_2010,romanczuk_mean-field_2012,mishra_collective_2012}. 

In this paper, we will derive a kinetic and a hydrodynamic description for self-propelled agents interacting via a selective attraction and repulsion interaction. A corresponding model was recently introduced to describe the onset of collective motion in insect swarms and is directly motivated by response of individual agents to looming visual stimuli and in particular the distinction of approaching and moving away objects\cite{guttal_cannibalism_2012,romanczuk_swarming_2012}. 

A fundamental difference of our model and the Vicsek model is its formulation in continuous time using stochastic differential equations. The interaction of individuals are modelled by superposition of (binary) interaction forces. 
This allows on the one hand a straight forward generalization of the model, e.g. towards variable speed of individuals, and on the other hand can be directly coarse-grained by deriving the Fokker-Planck equation for the corresponding probability density functions. The latter being the main focus of this work.
 
The model shows different phases including large scale collective motion, a disordered clustering phase and a nematic phase despite the absence of an explicit velocity-alignment interaction.

First, we will introduce the microscopic, individual-based model in terms of stochastic differential equations. Then we will proceed with the discussion of a kinetic description of the collective dynamics based on the nonlinear Fokker-Planck equation for the one-particle density function, which allows efficient numerical analysis of the stability of solutions of the Fokker-Planck equation in Fourier space. Further on, we will derive a hydrodynamic theory for self-propelled particles with selective attraction-repulsion interaction, which yields hydrodynamic equations in agreement with the theory by Toner and Tu. A direct comparison of the kinetic approach, which in principle can be considered up to arbitrary accuracy, to the hydrodynamic theory, which corresponds to a closure at the level of the second Fourier mode of the probability density function, reveals the range of validity of the hydrodynamic equations at large wavenumbers. Moreover, the analysis provides insights into the origin of unphysical divergences at large wavenumbers related to the approximations used namely the usage of Taylor polynomials. Finally, we compare the results of the kinetic and hydrodynamic theory with direct numerical simulations of the individual-based model.

	\section{Microscopic Model}

We consider $N$ self-propelled particles of mass $m=1$ in two spatial dimensions, so-called agents. Each individual moves at a constant speed $s_0$, thus the velocity vector of each agent is determined by its polar orientation angle $\varphi_i$. The equations of motion for the positions $\bi{r}_i$ and the polar orientation angles $\varphi_i$ read:
\eqnlabel{eq:eom_model}\numparts
\begin{eqnarray}
\frac{\rmd \bi{r}_i}{\rmd t} & = & s_0 \bi{ e}_{i,h}(t) = s_0 \! \left( \begin{array}{c} \cos\varphi_i(t) \\ \sin\varphi_i(t)  \\
\end{array}\right),\label{eq:eom_pos} \\ 
\frac{\rmd \varphi_i}{\rmd t} & = & \frac{1}{s_0}\left( F_{i,\varphi} + \sqrt{2 D_\varphi} \, \xi_i (t)\right) \label{eq:eom_angle}\, .
\end{eqnarray}
\endnumparts
Here, $F_{i,\varphi}=\bi{ F}_i \cdot \bi{ e}_{i,\varphi}$ is the projection of an effective social force vector $\bi{F}_i$ on the angular degree of freedom $\varphi_i$, which induces a turning behaviour of the focal individual due to interaction with others. $\bi{e}_{i,\varphi}=( -\sin\varphi_i, \cos\varphi_i)^T$ is the angular unit vector perpendicular to $\bi{e}_{i,h}$.

The second term within the brackets in \eref{eq:eom_angle} stands for random angular noise with intensity $D_\varphi$. $\xi_i(t)$ are independent, Gaussian random processes with vanishing mean and temporal $\delta$-correlations, i.e. $\mean{\xi_i(t) \xi_j(t+\tau)} = \delta_{ij} \, \delta(\tau)$ (Gaussian white noise).  

The total social force is given by a sum of three components:
\begin{eqnarray}
\bi{ F}_i & = \bi{ f}_{i,r}+\bi{ f}_{i,m} +\bi{ f}_{i,a} .
\end{eqnarray}
The first term represents a short-ranged repulsion 
allowing for finite sized agents. It reads 
\begin{eqnarray}\label{eq:epspp_repuls}
\bi{ f}_{i,r} & = -\sum_{j=1}^N \mu_r({r}_{ji}) \hat \bi{r}_{ji} \theta(l_c - r_{ji})
\end{eqnarray}
with $\mu_r({r}_{ji}) \geq 0 $ being a repulsive turning rate, which in general depends on distance $r_{ji} = \abs{\bi{r}_{ji}}=\abs{\bi{r}_j - \bi{r}_i}$ between two particles. We assume that this repulsive interaction is strictly short-ranged, i.e. it vanishes above a finite repulsive radius $l_c$, as indicated by the Heaviside (unit-step) function $\theta(x)$.

The other two forces read:
 \numparts
 \begin{eqnarray}
 \bi{ f}_{i,m} & =  \sum_{j=1}^N \mu_m({r}_{ji})  \hat \bi{ r}_{ji} 
 \theta(l_s -  r_{ji}) \theta(r_{ji} - l_c) \abs{ \tilde v_{ji} }  \theta(+\tilde v_{ji}), \label{eq:fm}\\
 \bi{ f}_{i,a} & =  \sum_{j=1}^N \mu_a({r}_{ji})  \hat \bi{ r}_{ji}  
 \theta(l_s -  r_{ji}) \theta(r_{ji} - l_c) \abs{ \tilde v_{ji} } \theta(-\tilde v_{ji}). \label{eq:fa}  
 \end{eqnarray}
 \endnumparts
These forces can be considered as a sum over binary interactions, which act always along the unit vector $\hat \bi{ r}_{ji}=(\bi{ r}_j-\bi{ r}_i)/r_{ji}$ pointing towards the center of mass of the neighbouring particle $j$. The corresponding response strengths $\mu_{a,m}(r_{ji})$ are distance dependent and may in general be both positive (attraction) and negative (repulsion). Furthermore, the overall response to other individuals is assumed to vanish above a finite sensory range $l_s$: $\mu_{a,m}(r_{ji}>l_s)=0$, whereby $l_s \geq l_c$.
\begin{figure}
\begin{center}
\centering\includegraphics[width=0.45\linewidth]{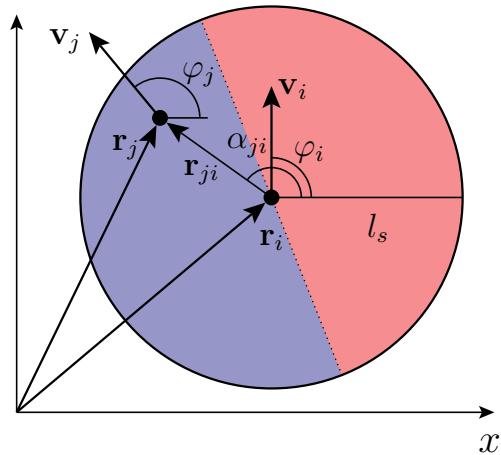}
\caption{Visualization of the spatial regions for approaching and receding self-propelled particles for a binary interaction of the $i$th particle with velocity vector ${\bf v}_i$ (heading angle $\varphi_i$), and a neighbouring particle $j$ with velocity ${\bf v}_j$ (heading angle $\varphi_j$) within its sensory range $l_s$.  The center of the circle corresponds to the position of the focal particle $i$. The dotted line, determined by the mean angle $(\varphi_i+\varphi_j)/2$, represents the border between the two distinct spatial regions (half-discs) corresponding to approaching and moving away of individual $j$. If the $j$th particle is located above the dotted line in the red region, the two particles are coming closer (approaching). Otherwise, if the neighbouring particle is located in the blue region, the two particles move away from each other (as shown in this example).  
\label{fig:scheme_angles}}
\end{center}
\end{figure}

The decisive factor in the distinction of the two forces is the sign of the relative velocity $\tilde v_{ji}$ defined by the temporal derivative of the distance ${\dot r}_{ji}$ between particles $i$ and $j$, and equals the projection of the velocity difference ${\bf v}_{ji}={\bf v}_j-{\bf v}_i$ of neighbour $j$ and the focal individual $i$ on the relative position unit vector $\hat{\bi{r}}_{ji}$: $\tilde v_{ji}={\bf v}_{ji}\cdot \hat{\bi{r}}_{ji}=\tilde v_{ij}$. Hence $\bi{f}_{i,a}$ is the response to approaching individuals characterized by a negative relative velocity $\tilde v_{ji} < 0$, whereas $\bi{ f}_{i,m}$ is the corresponding response to moving away (or receding) individuals characterized by positive relative velocity $\tilde v_{ji}>0$. Both force terms are proportional to the absolute value of the relative velocity, leading to stronger responses to faster approaching or receding individuals. 
Using the definition $\tilde{v}_{ji}=\dot{r}_{ji}$, one obtains the relative velocity
\begin{eqnarray}
  \label{eq:velo}
  \tilde v_{ji} = \tilde v_{ji}(\varphi_i,\varphi_j,\alpha_{ji})=2 {s}_0 \sin\left(\frac{\varphi_j+\varphi_i}{2}-\alpha_{ji}\right)\,\sin\left(\frac{\varphi_i-\varphi_j}{2}\right) 
\end{eqnarray}
as a function of the velocity angles $\varphi_{i}$, $\varphi_{j}$ and the angle $\alpha_{ji}$ of the distance vector $\bi{r}_{ji}$, with $\bi{r}_{ji}= r_{ji} \left ( \cos \alpha_{ji}, \sin \alpha_{ji} \right)$. From \eqnref{eq:velo}, one finds the two different spatial regions of approaching and receding particles, respectively, where the relative velocity has a different sign for fixed $\varphi_i$ and $\varphi_j$ in the interval $\varphi_i\le\varphi_j < \varphi_i + 2 \pi$:
\begin{eqnarray}
  \begin{array}{c c r c l c}
    \tilde{v}_{ji}  > 0   & \mbox{for} & \frac{\varphi_j+\varphi_i}{2}     & < \alpha_{ji} < & \frac{\varphi_j+\varphi_i}{2}+\pi & \mbox{moving away,}\\
    \tilde{v}_{ji}  < 0   & \mbox{for} & \frac{\varphi_j+\varphi_i}{2}+\pi & < \alpha_{ji} < & \frac{\varphi_j+\varphi_i}{2}+2\pi & \mbox{approaching.}
  \end{array} 
  \label{vrel}
\end{eqnarray}
Hence, a particle located in the half-sphere in clockwise direction from the mean angle $(\varphi_j+\varphi_i)/2$ approaches the focal particle $i$, whereas particles located in the other half-sphere anticlockwise from the mean angle are moving away (see figure \ref{fig:scheme_angles}). Please note that for $\varphi_j=\varphi_i$ the social force vanishes as $\tilde v_{ji}=0$ according to (\ref{eq:velo}).

\begin{figure}
\begin{center}
\includegraphics[width=0.95\textwidth]{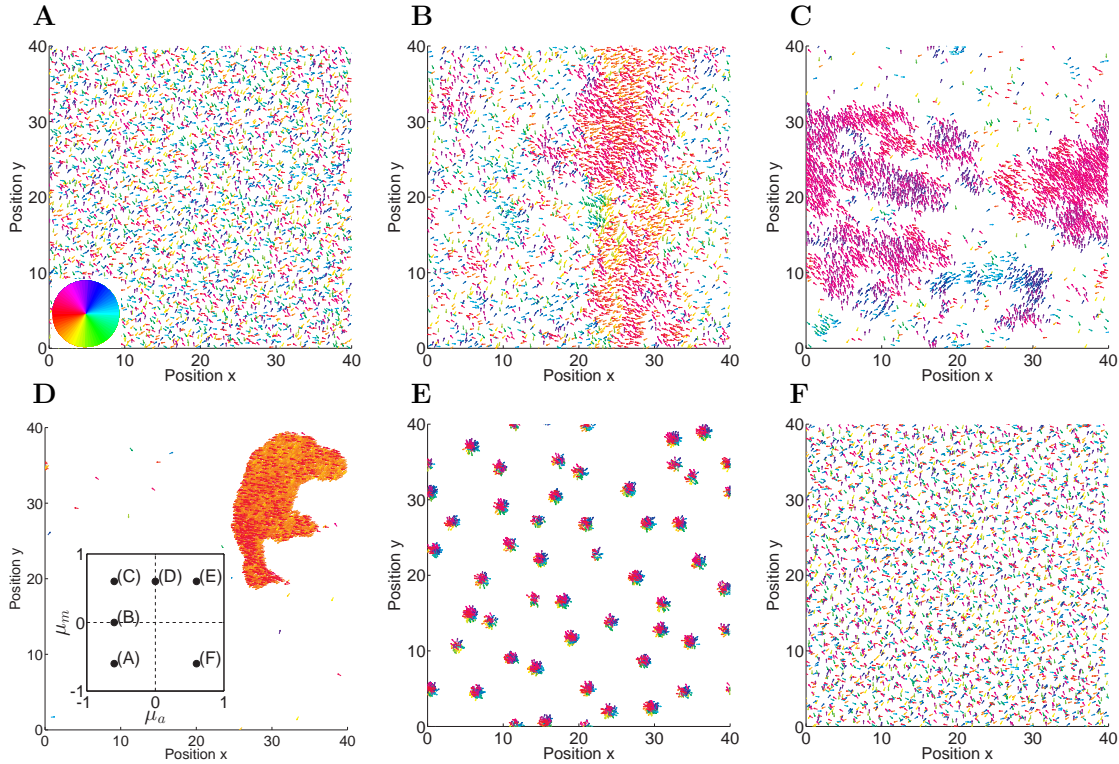}
\end{center}
\caption{
	Spatial snapshots of the microscopic model for different interaction strengths after the system relaxes towards a steady-state: (A) disordered, homogeneous state for $\mu_a=\mu_m=-0.6$ (pure repulsion), (B) diffuse collectively moving bands for $\mu_a=-0.6$, $\mu_m=0$ (repulsion from approaching individuals), (C) collectively moving bands for $\mu_a=-0.6$, $\mu_m=0.6$ (effective alignment), (D) dense, collectively moving cluster for $\mu_m=0.6$, $\mu_a=0$ (attraction to individuals moving away), (E) cluster state for $\mu_m=\mu_a=0.6$ (pure attraction), (F) disordered, homogeneous state for $\mu_m=-0.6$, $\mu_a=0.6$ (effective anti-alignment). The velocity vectors of individual particles are coloured corresponding to their polar direction of motion according to the inset of (A). The inset in (D) indicates the positions of the snapshots in the $(\mu_m,\mu_a)$-parameter plane. Other parameters: $N=4000$, $L=40$, $D_\varphi=0.02$, $l_s=1$, $l_c=0.2$, $\mu_r=5$, $s_0=0.25$.
\label{fig:snapshots}}
\end{figure}

Four different regions in the $(\mu_m,\mu_a)$-parameter space are distinguished \cite{romanczuk_swarming_2012}:
 \begin{itemize}
  \item pure attraction: $\mu_m>0$, $\mu_a > 0$; 
  \item effective alignment: $\mu_m>0$, $\mu_a < 0$, i.e. attraction to particles moving away, repulsion from particles coming closer;
  \item pure repulsion: $\mu_m<0$, $\mu_a < 0$;
  \item effective anti-alignment: $\mu_m<0$, $\mu_a > 0$, i.e. attraction to particles coming closer, repulsion from particles moving away.
 \end{itemize}
Some typical spatial snapshots obtained from individual-based simulations of \eref{eq:eom_model} are shown in figure \ref{fig:snapshots}. A schematic visualisation of the turning behavior of interacting agents in the different regimes is shown in \ref{app:visual}.

Please note that in previous publications \cite{guttal_cannibalism_2012,romanczuk_swarming_2012}, the ``effective alignment'' regime was referred to as ``escape and pursuit'', whereas ``effective anti-alignment'' was labelled ``head on head''. These previous labels had their origin in the context of heterogeneous agents and the respective response of individual agents to neighbors irrespective on their behavior.  Furthermore, an asymmetric alignment response for $\mu_m>0$, $\mu_a < 0$, yields in the present model the same qualitative dependence of the observed spatial patterns on the interaction strengths as in the original ``escape and pursuit'' model \cite{romanczuk_collective_2009}.  
Nevertheless, for self-propelled agents with constant speed, the above labels appear more appropriate.

One could argue, that in the effective alignment regime the macroscopic dynamics is essentially identical to other models, as for example the minimal Vicsek-model. However, for strong attraction to particles moving away in comparison to the repulsion to approaching particles, we observe the emergence of dense, collectively moving clusters, which do not appear in simple alignment models. 

Our model corresponds to the ones studied in \cite{guttal_cannibalism_2012,romanczuk_swarming_2012}, if the distance dependent interaction strengths $\mu_r(r_{ji})$ and $\mu_{a,m}(r_{ji})$ are assumed to be constant and the forces are rescaled by the number of particles within the interaction area of the focal particle. 

	\section{Kinetic Description}
	\label{sec:kinetic}

\subsection{Mean-Field Theory - Derivation and Analysis of a Nonlinear Fokker-Planck Equation}
\label{subsec:MFT}

In this section, we derive a kinetic description for the above individual-based model \eqnref{eq:eom_model}. For this purpose, we introduce the $N$-particle probability density function (PDF) 
 \begin{eqnarray}
   P_N(\bi{r}_1,\varphi_1; \bi{r}_2,\varphi_2; \ldots ; \bi{r}_N,\varphi_N ; t)\, ,
 \end{eqnarray}
which determines the probability to find particle $i$ at position $\bi{r}_i$ moving into the direction $\varphi_i$ ($i=1, 2, \ldots, N$) at time $t$. It is normalized with respect to integration over all positions and angles:
 \begin{eqnarray}
  \left [\; \prod_{j = 1 }^N \left ( \int \rmd^2 r_j \int_0^{2 \pi} \rmd \varphi_j \right) \right ] P_N(\bi{r}_1,\varphi_1; \bi{r}_2,\varphi_2; \ldots ; \bi{r}_N,\varphi_N ; t) = 1 \, . 
 \end{eqnarray}
In agreement with \eqnref{eq:eom_model}, we can write down the Fokker-Planck equation (FPE) for the dynamics of the PDF as follows.
 \begin{eqnarray}
   \label{eqn:FPE:fullN}
   \frac{\partial P_N}{\partial t} = - s_0 \sum_{i=1}^N \nabla_{\bi{r}_i} \cdot \left ( \bi{e}_{i,h} P_N \right) - \sum_{i=1}^N \frac{\partial}{\partial \varphi_i} \left ( \frac{ F_{i,\varphi} \, P_N }{s_0} \right) + \frac{D_{\varphi}}{s_0^2} \sum_{i=1}^N \frac{\partial^2 P_N}{\partial \varphi_i^2}
 \end{eqnarray}
From the linear FPE above one can derive an evolution equation for the marginal PDF
 \begin{eqnarray}
   \fl P(\bi{r}_i,\varphi_i,t) =  \left [ \; \prod_{j \neq i}^N \left ( \int \rmd^2 r_j \int_0^{2 \pi} \rmd \varphi_j \right) \right ] P_N(\bi{r}_1,\varphi_1; \bi{r}_2,\varphi_2; \ldots ; \bi{r}_N,\varphi_N ; t)
 \end{eqnarray}
by integrating \eref{eqn:FPE:fullN} over the degrees of freedom of particles $j \neq i$. In what follows, we assume that correlations between particles can be neglected.  Therefore, the $N$-particle PDF factorizes, i.e.
 \begin{eqnarray}
   \label{eqn:mean-field_assumpt}
   P_N(\bi{r}_1,\varphi_1; \bi{r}_2,\varphi_2; \ldots ; \bi{r}_N,\varphi_N ; t) = \prod_{i=1}^N P(\bi{r}_i,\varphi_i,t). 
 \end{eqnarray}
By this means, one obtains an effective one-particle description
 \begin{eqnarray}
   \label{eqn:fpe:one:particle}
   \frac{\partial P (\bi{r}_i,\varphi_i,t)}{\partial t} = - s_0 \nabla_{\bi{r}_i} \cdot \left (\bi{e}_{i,h} P \right) - \frac{N-1}{s_0} \frac{\partial }{\partial \varphi_i} \left ( F_{\varphi}\,  P \right) + \frac{D_\varphi}{s_0^2} \frac{\partial^2 P}{\partial \varphi_i^2} \, , 
 \end{eqnarray}
where the force $F_{\varphi}$ is given by the following integral over $P(\bi{r}_i,\varphi_i,t)$:
 \begin{eqnarray}
   \label{eqn:self:consistent:force}
   \fl F_{\varphi}&(\bi{r}_i,\varphi_i,t) = 2 s_0 \int_{\varphi_i}^{\varphi_i + 2 \pi} \mbox{d}\varphi_j \, \sin \left ( \frac{\varphi_i - \varphi_j}{2} \right)\int_{l_c}^{l_s} \mbox{d}r_{ji} \, r_{ji} \nonumber \\
 \fl & \left [ \mu_m(r_{ji}) \int_{\frac{\varphi_i + \varphi_j}{2}}^{\frac{\varphi_i + \varphi_j}{2} + \pi} \mbox{d}\alpha_{ji} \, \sin(\alpha_{ji} - \varphi_i) \sin \left ( \frac{\varphi_i + \varphi_j}{2} - \alpha_{ji} \right) P(\bi{r}_i+\bi{r}_{ji},\varphi_j,t) \right. \nonumber  \\
\fl  & \left. - \mu_a(r_{ji}) \int_{\frac{\varphi_i + \varphi_j}{2}+\pi}^{\frac{\varphi_i + \varphi_j}{2} + 2 \pi} \mbox{d}\alpha_{ji} \, \sin(\alpha_{ji} - \varphi_i) \sin \left ( \frac{\varphi_i + \varphi_j}{2} - \alpha_{ji} \right) P(\bi{r}_i+\bi{r}_{ji},\varphi_j,t) \right] \nonumber \\
\fl &- \int_{\varphi_i}^{\varphi_i+2\pi}\mbox{d}\varphi_j \int_{0}^{l_c} \mbox{d}r_{ji} \, r_{ji} \int_{0}^{2\pi} \mbox{d}\alpha_{ji} \, \mu_r(r_{ji}) \sin(\alpha_{ji} - \varphi_i) P(\bi{r}_i+\bi{r}_{ji},\varphi_j,t) \, .
 \end{eqnarray}
The factor $N-1$ in \eref{eqn:fpe:one:particle} arises, because \eref{eqn:FPE:fullN} was integrated over the positions and angles of $N-1$ identical particles. In other words, the focal particle can interact with $N-1$ identical neighbours. For the same reason, the particle index $i$ is omitted henceforward. 

The FPE \eqnref{eqn:fpe:one:particle}, which was derived from the linear FPE \eqnref{eqn:FPE:fullN} governing the dynamics of $P_N$, is nonlinear since the force terms $F_\varphi$ depend on $P(\bi{r},\varphi,t)$. In this sense, assumption \eqnref{eqn:mean-field_assumpt} is a mean-field approximation: a single particle is affected by a force due to its own PDF \eqnref{eqn:self:consistent:force}.  

By introducing the one-particle density function $p(\bi{r},\varphi,t)=N P(\bi{r},\varphi,t)$
we can eliminate the factor $(N-1) \approx N$ from \eqnref{eqn:fpe:one:particle}. Accordingly, $p(\bi{r},\varphi,t)$ is interpreted as particle density obeying the nonlinear Fokker-Planck equation
 \begin{eqnarray}
   \label{eqn:fpe:particle:density}
   \frac{\partial p (\bi{r},\varphi,t)}{\partial t} = - s_0 \nabla_{\bi{r}} \cdot \left (\bi{e}_h \, p \right) - \frac{\partial }{\partial \varphi} \frac{ F_{\varphi} (\bi{r},\varphi,t) \, p }{s_0} + \frac{D_\varphi}{s_0^2} \frac{\partial^2 p}{\partial \varphi^2} \, . 
 \end{eqnarray}

First, we assume a spatially homogeneous situation where $p(\bi{r},\varphi,t)= p(\varphi,t)$ holds. In this case, \eref{eqn:fpe:particle:density} is reduced to 
 \begin{eqnarray}
   \label{eqn:fpe:particle:density:spatially:homogeneous}
     \fl \frac{\partial p (\varphi,t)}{\partial t} = 
   - \frac{D_{\varphi}}{s_0^2} \frac{\partial }{\partial \varphi} \left [  \frac{\kappa}{\rho_0} \, \, \int_{\varphi}^{\varphi + 2 \pi} \rmd \varphi_j \, \sin (\varphi_j - \varphi) p(\varphi_j,t) \, p(\varphi,t) - \frac{\partial p}{\partial \varphi} \right] \, ,
 \end{eqnarray}
where the dimensionless coupling parameter 
 \begin{eqnarray}
   \label{eqn:dimless:kappa}
   \kappa = \frac{\pi \rho_0 s_0^2}{2 D_{\varphi}} \int_{l_c}^{l_s} \rmd r_{ji} \, r_{ji} \left ( \mu_m(r_{ji}) - \mu_a(r_{ji}) \right) 
 \end{eqnarray}
is introduced, where $\rho_0=N/L^2$ is the homogeneous particle density, with $L$ being the linear spatial dimension of the two-dimensional square-shaped system.

One obtains the same Fokker-Planck equation as \eref{eqn:fpe:particle:density:spatially:homogeneous} for self-propelled particles interacting via a velocity-alignment mechanism or globally coupled  Kuramoto oscillators \cite{shinomoto_phase_1986,sakaguchi_phase_1988}. Hence, the selective attraction-repulsion interaction reduces to effective velocity-alignment as considered in \cite{peruani_mean-field_2008} (continuum time form of the original Vicsek model \cite{vicsek_novel_1995}), if spatial inhomogeneities are neglected.
Furthermore, the effective alignment strength  is proportional to $\mu_m-\mu_a$ and may be negative, leading to anti-alignment as a consequence of repulsion from particles moving away and attraction to particles coming closer, respectively (see also figure \ref{fig:scheme}(D)).  

The spatially homogeneous, time-independent particle density
 \begin{eqnarray}
   \label{eqn:spatially:homogeneous:ordered:solution}
   p^{(\Phi)}(\varphi) = \frac{\rho_0}{2 \pi} \, \frac{\exp \left ( \kappa  \Phi \cos (\varphi - \varphi_0) \right)}{I_0(\kappa \Phi)} 
 \end{eqnarray}
is a solution to both nonlinear Fokker-Planck equations \eref{eqn:fpe:particle:density} and \eref{eqn:fpe:particle:density:spatially:homogeneous}, as can be easily shown by inserting \eref{eqn:spatially:homogeneous:ordered:solution} into \eref{eqn:fpe:particle:density:spatially:homogeneous}. 
$I_\nu(x)$ denotes the modified Bessel-function of the first kind. Without loss of generality, we choose the frame of reference such that the direction of collective motion $\varphi_0$ defines the x-coordinate, i.e. $\varphi_0 = 0$. 
From the definition of the polar order parameter 
 \begin{eqnarray}
   \Phi & = & \abs{\mean{e^{i \varphi}} } = \abs{\int_0^{2\pi} d\varphi \, p^{(\Phi)}(\varphi) e^{i \varphi}}  
 \end{eqnarray}
 we can determine the order parameter by solving  the transcendental equation 
 \begin{eqnarray}
   \label{eqn:bestimmungsgleichung:order:parameter}
     \Phi   & = & \frac{I_1(\kappa  \Phi)}{I_0(\kappa  \Phi)} \ .
 \end{eqnarray}

\Eref{eqn:bestimmungsgleichung:order:parameter} is always fulfilled for $\Phi = 0$. In that case, \eref{eqn:spatially:homogeneous:ordered:solution} yields the uniform distribution (spatially homogeneous, disordered state), i.e.
 \begin{eqnarray}
   \label{eqn:spatially:angular:homogeneous:solution}
   p^{(0)} = \frac{\rho_0}{2 \pi} \, .
 \end{eqnarray}
If $\kappa > 2$ (high effective alignment strength $\mu_m - \mu_a$ and low noise $D_{\varphi}$, respectively), there is a second nontrivial solution $\Phi > 0$ to \eref{eqn:bestimmungsgleichung:order:parameter} describing a spatially homogeneous state with polar order (collective motion; swarming phase). In this parameter region, the distribution $p^{(\Phi)}$ describing polar order ($\Phi > 0$) is stable with respect to spatially homogeneous perturbations $\delta p = \delta p(\varphi,t)$, whereas the homogeneous distribution $p^{(0)}$ is stable for $\kappa < 2$ \cite{shinomoto_phase_1986,sakaguchi_phase_1988}. 

The critical line $\kappa = 2$ for the order-disorder transition is translated to microscopic model parameters as follows: 
 \begin{eqnarray}
     \label{eqn:crit:line:coll:mot:from:distribution}
     \int_{l_c}^{l_s} \rmd r_{ji} \, r_{ji} \left( \mu_m(r_{ji}) - \mu_a(r_{ji}) \right) = \frac{4 D_{\varphi}}{\pi \rho_0 s_0^2} \, \mbox{. }
 \end{eqnarray}
 The above is a generalization of the previous result obtained in  \cite{romanczuk_swarming_2012}.
Since the order-disorder transition described by \eref{eqn:bestimmungsgleichung:order:parameter} is continuous, the right hand side of \eref{eqn:bestimmungsgleichung:order:parameter} can be expanded in a Taylor series so that the dependence $\Phi=\Phi(\kappa)$ is known close to the critical point $\kappa_c = 2$ analytically: 
 \begin{eqnarray}
      \label{eqn:scaling:crit:point:FPE}
      \Phi(\kappa) \simeq \sqrt{\kappa - 2} + \mathcal{O}\left ((k-2)^{3/2} \right )      \, .
 \end{eqnarray}
The stability analysis of the spatially homogeneous solutions $p^{(\Phi)}$ with respect to arbitrary perturbations $\delta p = \delta p(\bi{r},\varphi,t)$ is carried out in the next section.  

\subsection{Stability Analysis in Fourier Space}
\label{sec:stability:in:fourier}

In this section it is shown, how the stability of solutions of the nonlinear Fokker-Planck equation \eref{eqn:fpe:particle:density} can be analyzed in Fourier space. A similar approach was used by Chou \etal in order to achieve a kinetic theory for self-propelled particles with metric-free interactions \cite{chou_ihle_2012}. 

In order to analyze the Fokker-Planck equation \eref{eqn:fpe:particle:density} analytically and especially to solve the integral \eqnref{eqn:self:consistent:force}, it is convenient to work in Fourier space with respect to the spatial coordinates $\bi{r}$ and the angular variable $\varphi$. The dynamics of the Fourier coefficients
 \begin{eqnarray}
   \fourier{g}{n}(\bi{k},t) = \int \rmd^2 r \int_0^{2 \pi} \rmd \varphi \, p(\bi{r},\varphi,t) e^{i \bi{k}\cdot\bi{r} + i n \varphi }
 \end{eqnarray}
reads
 \begin{eqnarray}
   \label{eqn:hierarchie:fourier_complete:allg}
   \fl \frac{\partial \fourier{g}{n}(\bi{k},t)}{\partial t} - \frac{i k_x s_0}{2} \left ( \fourier{g}{n+1} + \fourier{g}{n-1} \right) - \frac{k_y s_0}{2} \left (\fourier{g}{n+1} - \fourier{g}{n-1} \right) = -n^2 \frac{D_{\varphi}}{s_0^2} \fourier{g}{n} \nonumber \\
   + \frac{i n }{(2 \pi)^3 s_0} \sum_{j=-\infty}^{\infty} \int \rmd^2 q \, \fourier{g}{j}(\bi{q},t) \fourier{K}{n-j}(\bi{k}-\bi{q},t) \, ,
 \end{eqnarray}
where the Fourier coefficients of the force \eref{eqn:self:consistent:force} are defined as
 \begin{eqnarray}
   \fourier{K}{n}(\bi{k},t) = \int \rmd^2 r \int_0^{2 \pi} \rmd \varphi \, F_{\varphi}(\bi{r},\varphi,t) e^{i \bi{k}\cdot\bi{r} + i n \varphi} \, .
 \end{eqnarray}
Obviously, the Fokker-Planck equation turns into an infinite hierarchy of equations \eqnref{eqn:hierarchie:fourier_complete:allg} in Fourier space. 

For the stability analysis, the Fourier coefficients $\fourier{K}{n}(\bi{k},t)$ of the force \eqnref{eqn:self:consistent:force} are required. Due to the fact, that the force $F_{\varphi}(\bi{r},\varphi,t)$ depends on the integral over $p(\bi{r},\varphi,t)$ \eref{eqn:self:consistent:force}, the Fourier coefficients $\fourier{K}{n}(\bi{k},t)$ of the force can be written as a linear combination of the Fourier coefficients $\fourier{g}{r}(\bi{k},t)$
 \begin{eqnarray}
   \label{eqn:fourierCoefficients:n:k}
   \fourier{K}{n}(\bi{k},t) = \sum_{r=-\infty}^{\infty} \tilde{K}_{n,r}(\bi{k}) \, \fourier{g}{r}(\bi{k},t)
 \end{eqnarray}
using an infinite dimensional matrix $\tilde{K}_{n,r}(\bi{k})$. Details on the calculation of the matrix elements are given in \ref{sec:app:deriv:force:matrix}. Consequently, the dynamics of the Fourier coefficients $\fourier{g}{n}(\bi{k},t)$ read
 \begin{eqnarray}
   \fl \frac{\partial \fourier{g}{n}(\bi{k},t)}{\partial t} - \frac{i k_x s_0}{2} \left ( \fourier{g}{n+1} + \fourier{g}{n-1} \right) - \frac{k_y s_0}{2} \left (\fourier{g}{n+1} - \fourier{g}{n-1} \right) = -n^2 \frac{D_{\varphi}}{s_0^2} \fourier{g}{n} \nonumber \\
   + \frac{i n }{(2 \pi)^3 s_0} \sum_{j=-\infty}^{\infty} \sum_{r=-\infty}^{\infty} \int \rmd^2 q \, \fourier{g}{j}(\bi{q},t) \fourier{g}{r}(\bi{k} - \bi{q},t) \tilde{K}_{n-j,r}(\bi{k} - \bi{q}) \, .
 \end{eqnarray}

Let $\fourier{g}{n}^{(0)}(\bi{k},t)$ be a solution of the equation above. The dynamics of a small perturbation $\delta \fourier{g}{n}(\bi{k},t) = \fourier{g}{n}(\bi{k},t) - \fourier{g}{n}^{(0)}(\bi{k},t)$ is determined by the linearized equation
 \begin{eqnarray}
   \label{eqn:linear:fpe:in:fourier}
   \fl \frac{\partial \delta \fourier{g}{n}(\bi{k},t)}{\partial t} - \frac{i k_x s_0}{2} \left ( \delta \fourier{g}{n+1} + \delta \fourier{g}{n-1} \right) - \frac{k_y s_0}{2} \left (\delta \fourier{g}{n+1} - \delta \fourier{g}{n-1} \right) = -n^2 \frac{D_{\varphi}}{s_0^2} \delta \fourier{g}{n} + \frac{i n }{(2 \pi)^3 s_0} \nonumber \\
   \fl \sum_{j=-\infty}^{\infty} \sum_{r=-\infty}^{\infty} \int \rmd^2 q \, \left (\delta \fourier{g}{j}(\bi{q},t) \fourier{g}{r}^{(0)}(\bi{k}-\bi{q},t) + \fourier{g}{j}^{(0)}(\bi{q},t) \delta \fourier{g}{r}(\bi{k}-\bi{q},t) \right ) \tilde{K}_{n-j,r}(\bi{k} - \bi{q})  .
 \end{eqnarray}
The stability analysis of the solutions 
 \begin{eqnarray}
   p^{(\Phi)}(\varphi) = \frac{\rho_0}{2 \pi} \, \frac{\exp \left ( \kappa \Phi \cos \varphi \right)}{I_0(\kappa  \Phi)} 
 \end{eqnarray}
requires the corresponding Fourier coefficients 
 \begin{eqnarray}
   \fourier{g}{n}(\bi{k}; \Phi) &=(2 \pi)^2 \rho_0 \, \frac{I_n(\kappa \Phi)}{I_0(\kappa  \Phi)} \, \delta (\bi{k}) \, .
 \end{eqnarray}
Inserting those in \eref{eqn:linear:fpe:in:fourier} yields the following linearized ordinary differential equations
 \begin{eqnarray}
   \label{eqn:LODE:in:fourier:space}
   \frac{\partial \delta \fourier{g}{n}(\bi{k},t)}{\partial t} = \sum_{r=-\infty}^{\infty} \tilde{M}_{n,r}(\bi{k}) \, \delta \fourier{g}{r}(\bi{k},t) 
 \end{eqnarray}
with the matrix
 \begin{eqnarray}
   \label{eqn:LODE:in:fourier:space:matrix}    
   \fl \tilde{M}_{n,r}(\bi{k}) =& \frac{i k_x s_0}{2} \left ( \delta_{n+1,r} + \delta_{n-1,r} \right) + \frac{k_y s_0}{2} \left ( \delta_{n+1,r} - \delta_{n-1,r} \right) - n^2 \frac{D_{\varphi}}{s_0^2} \, \delta_{n,r} \nonumber \\ 
   \fl & + \frac{i n \rho_0}{2 \pi s_0} \sum_{j=-\infty}^{\infty} \frac{I_j(\kappa  \Phi)}{I_0(\kappa \Phi)} \left ( \tilde{K}_{n-r,j}(0) + \tilde{K}_{n-j,r}(\bi{k}) \right) \, .
 \end{eqnarray}
The eigenvalues $\sigma$ of the matrix above determine the stability of $p^{(\Phi)}$. In order to calculate the eigenvalues numerically as a function of the wavevector $\bi{k}$ (dispersion relation $\sigma(\bi{k})$), it is necessary to find an appropriate closure of the infinite dimensional system of linear equations. Here we assume, that a critical $\tilde{n}>0$ exists, such that $\delta \fourier{g}{n} = 0$ for $\abs{n}>\tilde{n}$ holds approximatively (cf. \cite{chou_ihle_2012}). That assumption is reasonable, because the $n$th Fourier coefficient is strongly damped, since
 \begin{eqnarray}
   \frac{\partial \delta \fourier{g}{n}(\bi{k},t)}{\partial t} \propto - n^2 \frac{D_{\varphi}}{s_0^2} \, \delta \fourier{g}{n}(\bi{k},t) \, .
 \end{eqnarray}
For the numerical analysis, $\tilde{n}=50$ is assumed. Using this approximation, the stability of the spatially homogeneous states can be studied. Please note, that only two approximations were used: First, correlations between particles are neglected \eqnref{eqn:mean-field_assumpt} (mean-field); Second, the truncation of the hierarchy of equations \eqnref{eqn:LODE:in:fourier:space}. In principle, it is possible to analyze the stability for arbitrary wavevectors $\bi{k}$, notably, there is no restriction to small wavenumbers $k$. The procedure described above implies arbitrary directions of perturbations (with respect to the direction of collective motion) of the homogeneous distribution $p^{(\Phi)}$. However, the wavevector $\bi{k}=(k,0)^T$ is chosen parallel with respect to the direction of collective motion. 

\begin{figure}
 \begin{center}
  \includegraphics[width=0.5\textwidth]{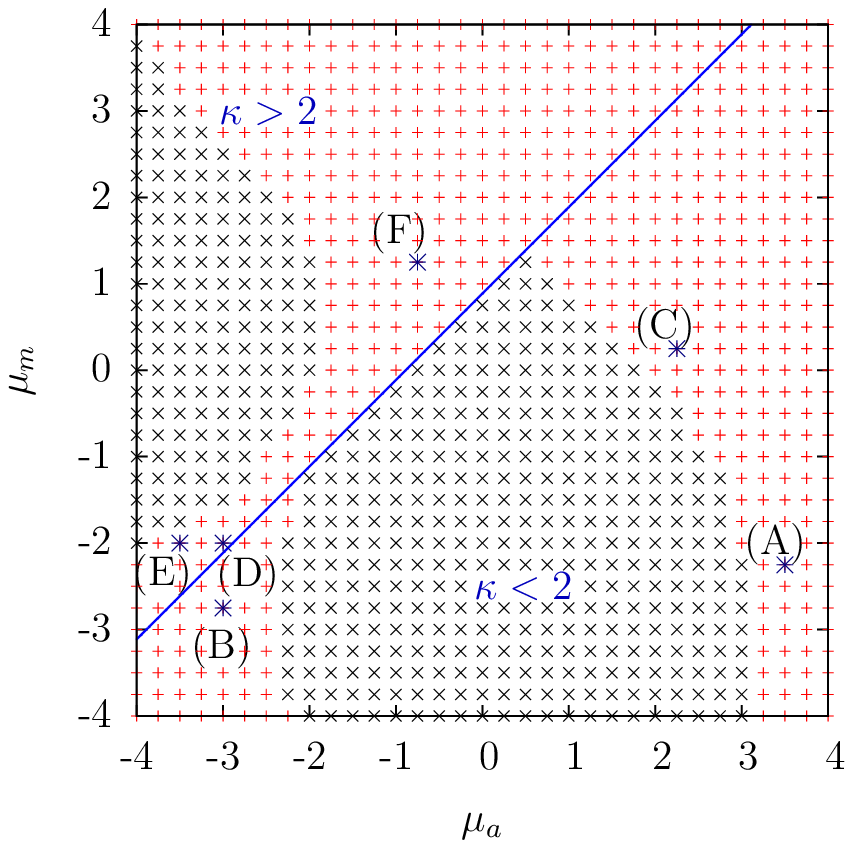}
  \caption{Stability of the solutions $p^{(\Phi)}$ in the $(\mu_m,\mu_a)$-parameter space as predicted by the kinetic theory (truncation of \eref{eqn:LODE:in:fourier:space:matrix} at $\tilde{n}=50$): black crosses indicate stable solutions, red plus signs indicate instabilities. The blue (solid) line corresponds to the critical line $\kappa =2$ \eref{eqn:crit:line:coll:mot:from:distribution}. Collective motion is possible for $\kappa > 2$. The dispersion relation for the specific points (A)-(F) is given in figure \ref{fig:larg:eigenval:kin:theo}. Whereas an interval of wavenumbers becomes unstable in (A) and (B), long-wavelength instabilities are found in (C)-(F). Parameters: $D_{\varphi} = 0.5$; $\rho_0 = 1.5$; $l_c = 0.2$; $l_s = 1$; $s_0 = 1$; $\mu_r = 1$; $\mu_a$ and $\mu_m$ piecewise constant. }
  \label{fig:phase:space:stability:ana:kin:theo}
 \end{center}
\end{figure}

\begin{figure}
 \begin{center}
	\begin{minipage}{0.48\textwidth}
 	\begin{center}
  	\includegraphics[width=\textwidth]{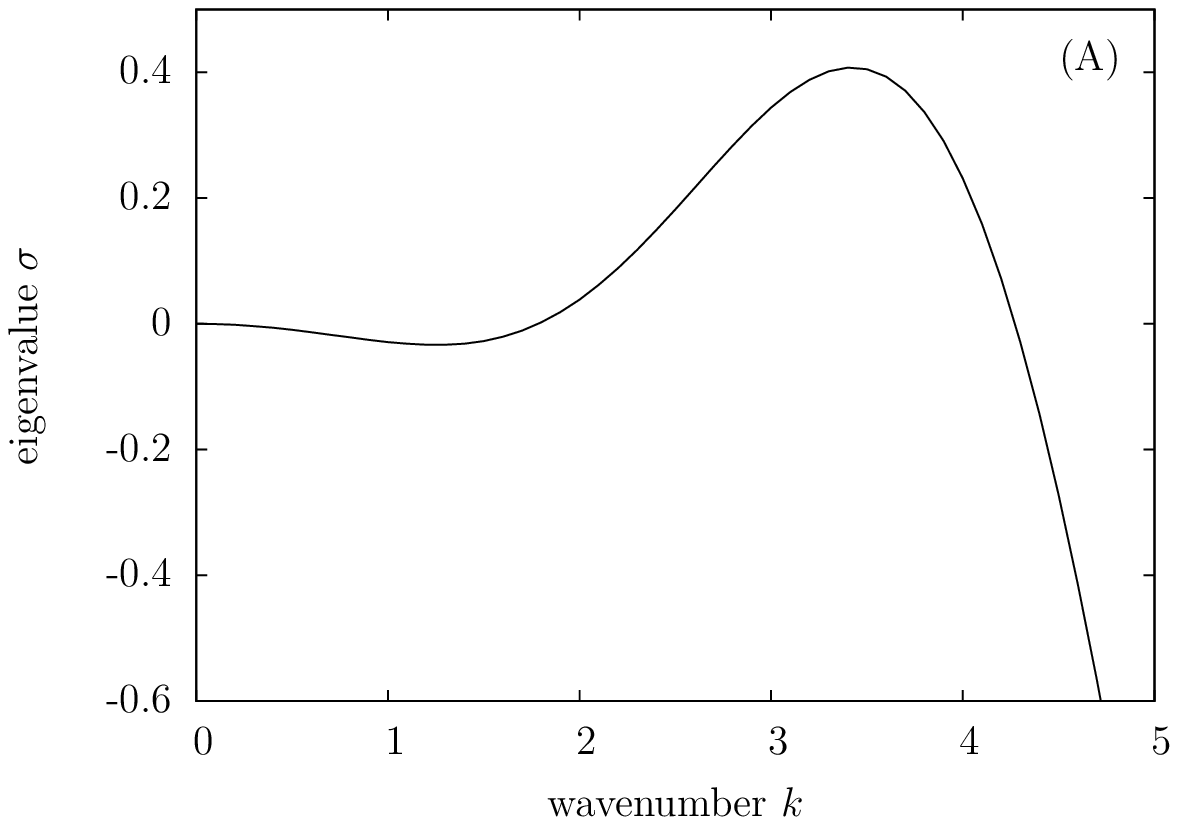}
 	\end{center}
	\end{minipage}
	\begin{minipage}{0.48\textwidth}
 	\begin{center}
  	\includegraphics[width=\textwidth]{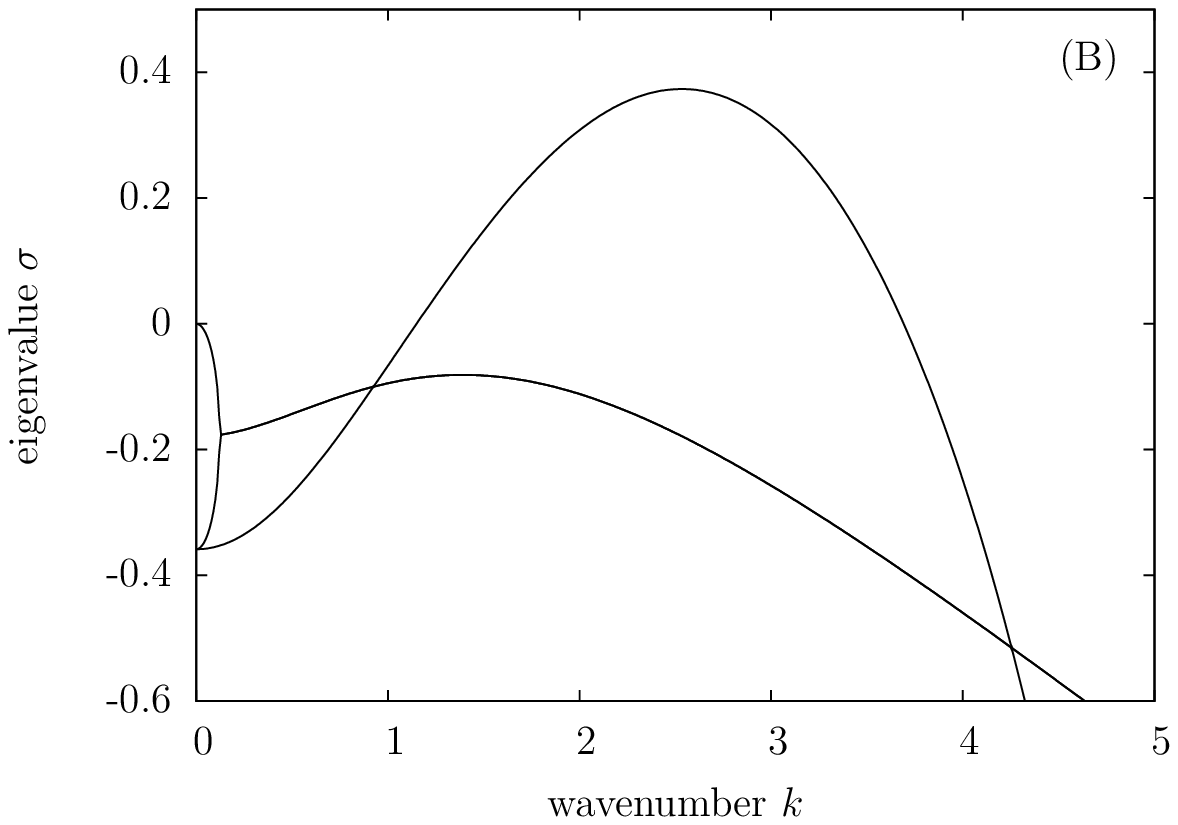}
 	\end{center}
	\end{minipage} 
	\begin{minipage}{0.48\textwidth}
 	\begin{center}
  	\includegraphics[width=\textwidth]{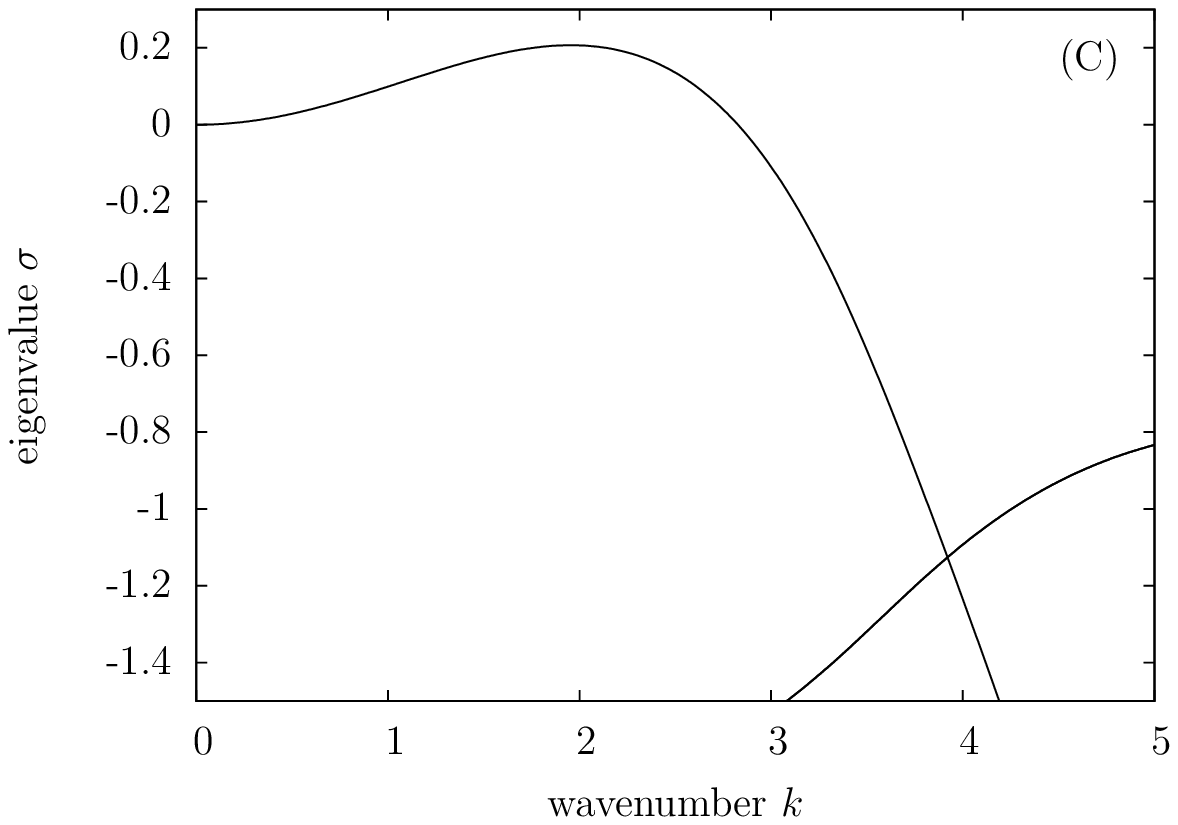}
 	\end{center}
	\end{minipage}
	\begin{minipage}{0.48\textwidth}
 	\begin{center}
  	\includegraphics[width=\textwidth]{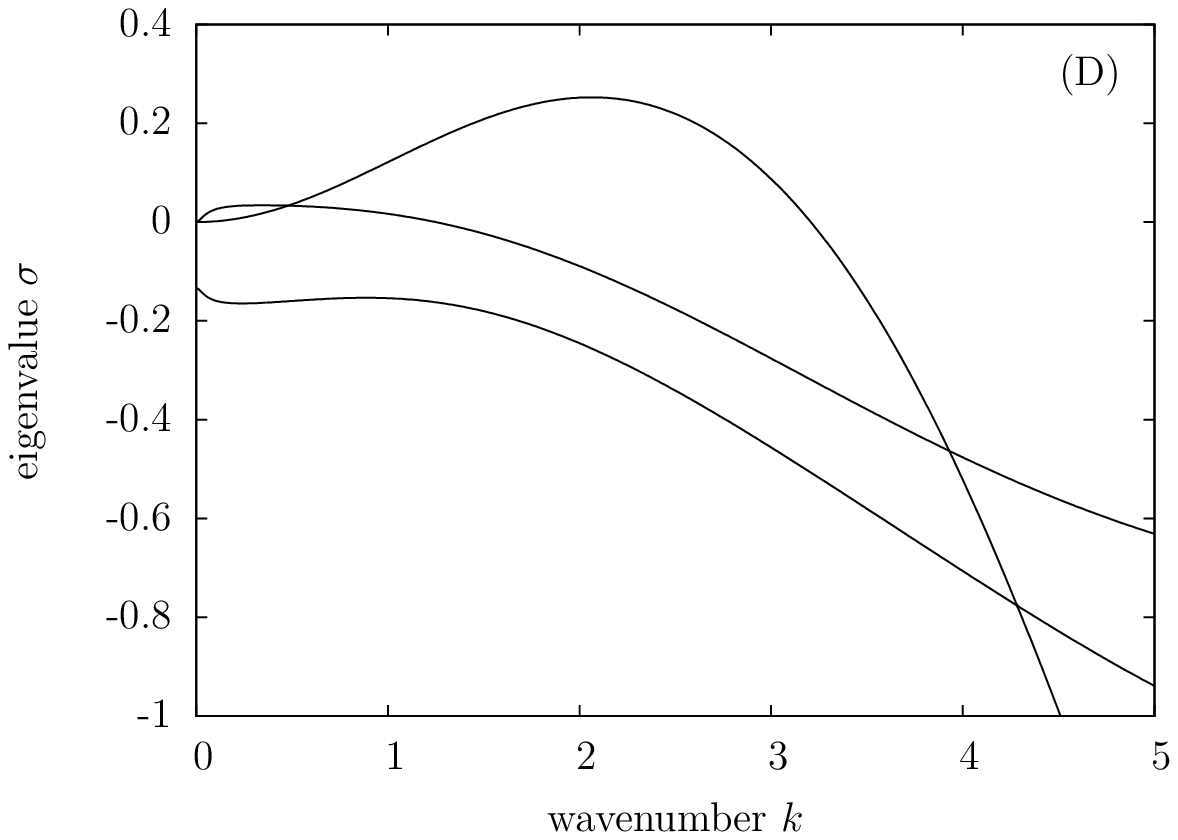}
 	\end{center}
	\end{minipage}
	\begin{minipage}{0.48\textwidth}
 	\begin{center}
  	\includegraphics[width=\textwidth]{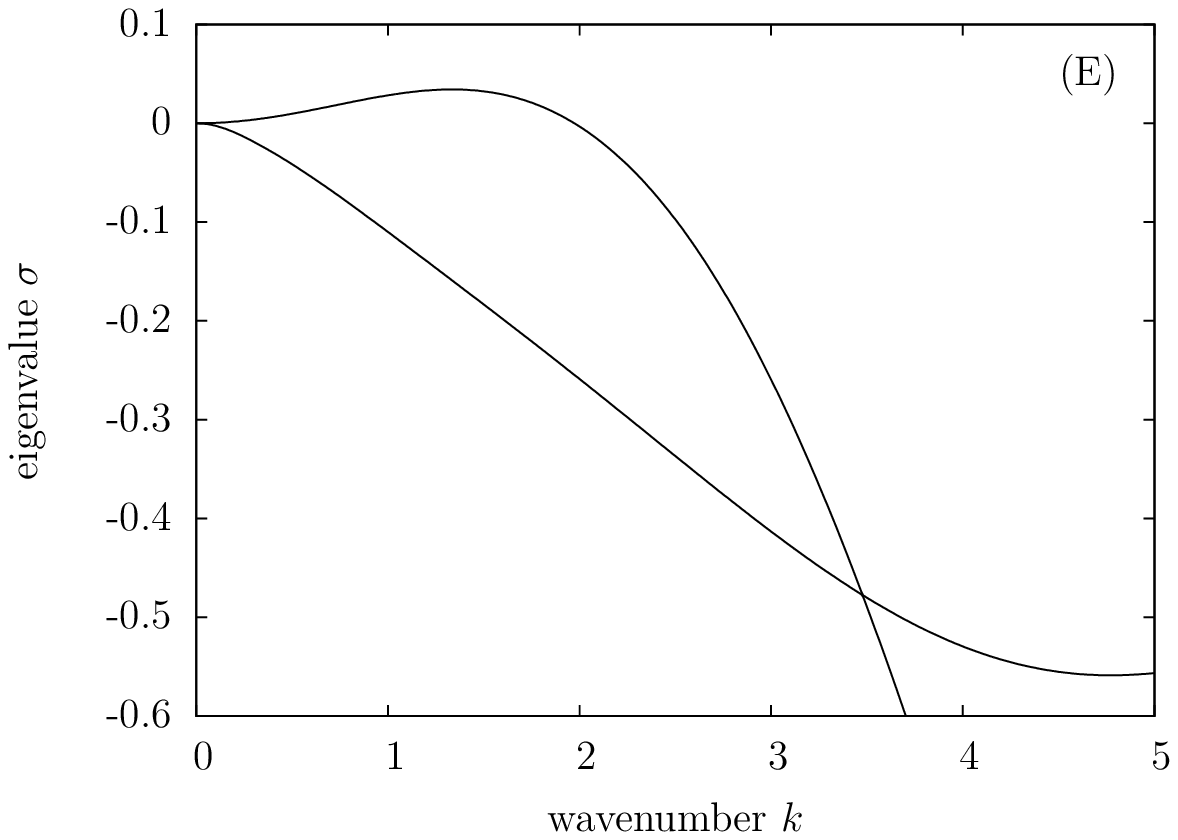}
 	\end{center}
	\end{minipage}
	\begin{minipage}{0.48\textwidth}
 	\begin{center}
  	\includegraphics[width=\textwidth]{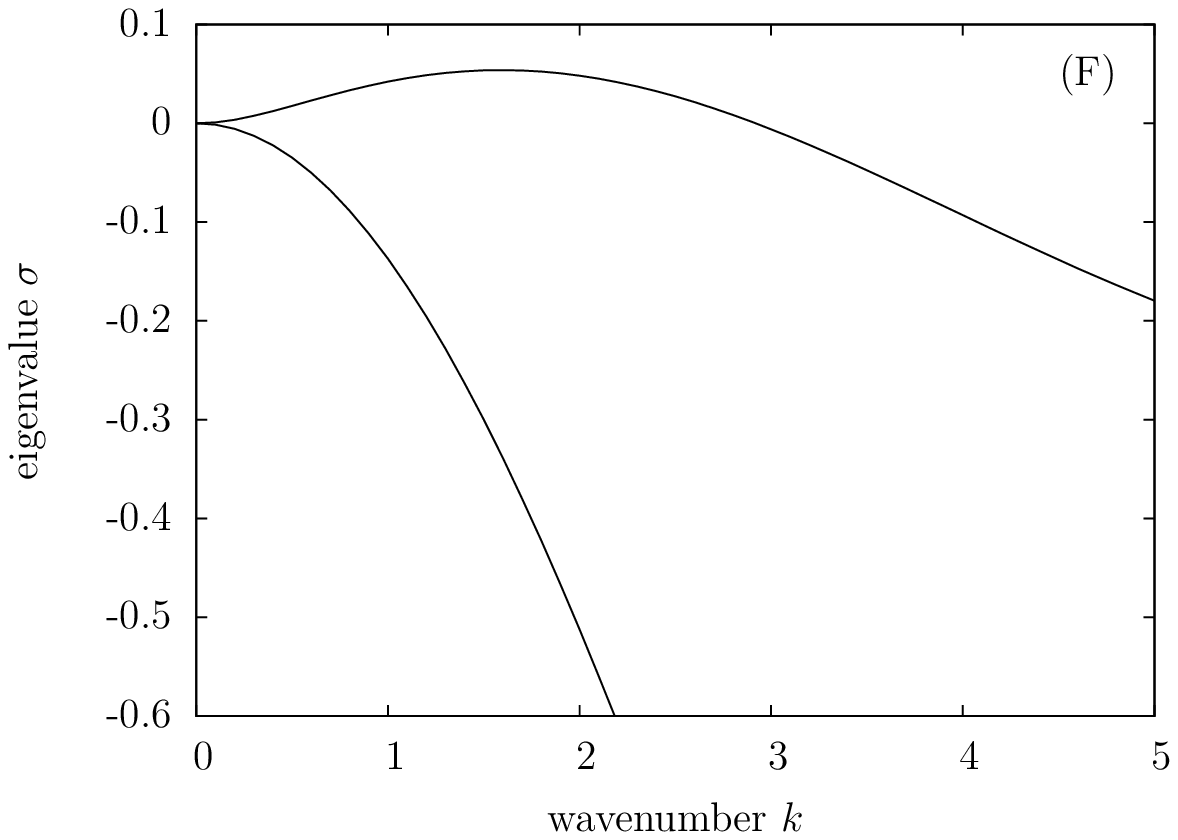}
 	\end{center}
	\end{minipage}
 \caption{Dispersion relation $\sigma(\bi{k}=(k,0))$ for specific points indicated in figure \ref{fig:phase:space:stability:ana:kin:theo}. Whereas an interval of wavenumbers becomes unstable in (A) and (B), long-wavelength instabilities are found in (C)-(F). Parameters: (A) $\mu_a = 3.5$, $\mu_m = -2.25$; (B) $\mu_a = -3$, $\mu_m = -2.75$; (C) $\mu_a = 2.25$, $\mu_m = 0.25$; (D) $\mu_a = -3$, $\mu_m = -2$; (E) $\mu_a = -3.5$, $\mu_m = -2$; (F) $\mu_a = -0.75$, $\mu_m = 1.25$. Other parameters: $D_{\varphi} = 0.5$; $\rho_0 = 1.5$; $l_c = 0.2$; $l_s = 1$; $s_0 = 1$; $\mu_r = 1$; $\mu_a$ and $\mu_m$ piecewise constant. }
  \label{fig:larg:eigenval:kin:theo}
 \end{center}
\end{figure}

The illustrated method allows to explore the structure of the phase space of the dynamical system. In particular, we are interested in the stability of the solutions $p^{(\Phi)}$ in the $(\mu_m,\mu_a)$-parameter space. For $\kappa < 2$, the stability of $p^{(0)}=\rho_0/(2 \pi)$ is investigated ($\Phi = 0$), whereas the solution $p^{(\Phi)}(\varphi)$ ($\Phi>0$) describing polar order is inserted in \eref{eqn:LODE:in:fourier:space:matrix} for $\kappa>2$. Figures \ref{fig:phase:space:stability:ana:kin:theo} and \ref{fig:larg:eigenval:kin:theo} show the results of the stability analysis in the $(\mu_m,\mu_a)$-parameter space and the dispersion relations for specific points (named (A)-(F) in figure \ref{fig:phase:space:stability:ana:kin:theo} and figure \ref{fig:larg:eigenval:kin:theo}), respectively. 

The kinetic approach reveals instabilities nearby the points (A) and (B). In these parameter regions, an interval of wavenumbers becomes unstable (cf. figure \ref{fig:larg:eigenval:kin:theo}). In (A), nematic filaments are found in the individual based simulations (see section \ref{sec:simulations}). However, a long-wavelength instability of the spatially homogeneous, disordered state $p^{(0)}$ is found in (C). According to the individual based simulations, a clustering phase emerges nearby the point (C) (cf. figure \ref{fig:snapshots}E for a snapshot of a related simulation). 

Close to the order-disorder transition line in the regime of collective motion ($\kappa > 2$), a long-wavelength instability emerges as can be seen in figure \ref{fig:larg:eigenval:kin:theo}(D). According to figures \ref{fig:larg:eigenval:kin:theo}(D)-(F), two hydrodynamic modes exist, which satisfy $\sigma(\bi{k}\rightarrow 0) \rightarrow 0$, corresponding to transversal and longitudinal perturbations with respect to the direction of collective motion. The destabilization of the spatially homogeneous, ordered state $p^{(\Phi)}$ is always determined by these hydrodynamic modes, leading to long-wavelength instabilities. 

Furthermore, there exists a region in the parameter space, where $p^{(\Phi)}$ is stable against spatially dependent perturbations. According to figure \ref{fig:phase:space:stability:ana:kin:theo}, this region is approximately bounded by the critical line $\kappa = 2$ \eref{eqn:crit:line:coll:mot:from:distribution} and the secondary diagonal $\mu_m = - \mu_a$, which is equally confirmed by the individual based simulation (see section \ref{sec:simulations}). 

In summary, it can be stated that the kinetic approach enables the prediction of the structure of the phase space as well as the dispersion relation for any microscopic model parameters. However, the analysis involves numerical methods. Unfortunately, it is not possible to deduce further analytical criteria for the stability of the solutions considered. For that purpose, additional assumptions are necessary, namely the restriction to the lowest Fourier coefficients and small wavenumbers, as shown in section \ref{sec:hydrodynamic}. The hydrodynamic theory supplies analytical criteria for the stability of the solutions considered, such as critical microscopic model parameters. 

In this context, we would like to emphasize one difficulty which occurs if the focus is put on the dynamics of the lowest Fourier coefficients on large length scales, i.e. small wavenumbers $k$: The matrix elements $\tilde{K}_{n,r}(\bi{k})$ involve Bessel functions of the first kind of $k$ (see \ref{sec:app:deriv:force:matrix} and \cite{chou_ihle_2012}), which are oscillating functions with alternating Taylor coefficients \cite{bronstein}. The approximation
 \begin{eqnarray}
   J_{\nu,N}(k x) \approx \sum_{m=0}^N \frac{(-1)^m}{m! \, \Gamma(m+\nu+1)} {\left( \frac{k x}{2} \right)}^{2m+\nu}
 \end{eqnarray}
will only be valid in the vicinity of $k = 0$. For large $k$, $J_{\nu,N}(k x)$ tends to either plus or minus infinity, but never to zero. This issue will be important for the derivation of hydrodynamic equations in the next section and is restricting their validity to small wavenumbers $k$ and large length scales, respectively. 

\section{Derivation of Hydrodynamic Equations}
\label{sec:hydrodynamic}

\subsection{Expansion of the Fokker-Planck Equation in the Angular Fourier Domain}

In this section, a hydrodynamic description of the many-particle system is derived directly from the one-particle FPE \eref{eqn:fpe:particle:density}. For this purpose, it is convenient to work in Fourier space with respect to the angular variable $\varphi$. Both, the particle density $p(\bi{r},\varphi,t)$ and the force $F_{\varphi}(\bi{r},\varphi,t)$ \eqnref{eqn:self:consistent:force} are $2\pi$-periodic functions in $\varphi$. Hence, they can be expanded in a Fourier series as follows: 
 \numparts
  \begin{eqnarray}
    p(\bi{r},\varphi,t)           &= \frac{1}{2 \pi} \sum_{n=-\infty}^{\infty} \fourier{f}{n} (\bi{r},t) \, e^{- i n \varphi} \, \mbox{, } \label{eqn:angular:fourierseries:OPPDF} \\
    F_{\varphi}(\bi{r},\varphi,t) &= \frac{1}{2 \pi} \sum_{n=-\infty}^{\infty} \fourier{Q}{n} (\bi{r},t) \, e^{- i n \varphi} \, \mbox{. } 
  \end{eqnarray}
 \endnumparts
The Fokker-Planck equation in Fourier space is obtained by multiplying \eqnref{eqn:fpe:particle:density} with $e^{i n \varphi}$ and integration over $\varphi \in \left [0,2 \pi \right ]$:
 \begin{eqnarray}
    \label{eqn:hierarchie:fourier_varphi:allg}
    \frac{\partial \fourier{f}{n}}{\partial t} + s_0 \left ( \nabcp \fourier{f}{n-1} + \nabcp^{*} \fourier{f}{n+1} \right) = \frac{1}{2 \pi} \frac{i n}{s_0} \sum_{j=-\infty}^{\infty} \fourier{f}{j} \, \fourier{Q}{n-j} - n^2 \frac{D_{\varphi}}{s_0^2} \fourier{f}{n} \, \mbox{. }  
 \end{eqnarray}
In \eqnref{eqn:hierarchie:fourier_varphi:allg}, the complex derivative $\nabcp = 0.5\left (\partial_x + i \partial_y \right)$ was introduced. The Fourier coefficients
 \begin{eqnarray}
   \fourier{f}{n}(\bi{r},t) = \int_{0}^{2 \pi} \rmd \varphi \, p(\bi{r},\varphi,t) e^{i n \varphi } = \rho(\bi{r},t) \mean{ e^{i n \varphi }} 
 \end{eqnarray}
obey the symmetry $\fourier{f}{n}^{*}(\bi{r},t) = \fourier{f}{-n}(\bi{r},t)$, since $p(\bi{r},\varphi,t)$ is real-valued. The lowest Fourier coefficients are related to macroscopic physical quantities, namely the marginal particle density $\rho(\bi{r},t)$ and also the momentum field $\bi{w}(\bi{r},t)=(w_x,w_y)$ via
 \eqnlabel{eqn:complex:numbers:real:quantities}
 \numparts
  \begin{eqnarray}
   \rho(\bi{r},t) &= \fourier{f}{0}(\bi{r},t) \, , \\
   w_x(\bi{r},t)  &= \frac{s_0}{2} \left ( \fourier{f}{1}(\bi{r},t) + \fourier{f}{-1}(\bi{r},t)  \right) \, , \\
   w_y(\bi{r},t)  &= \frac{s_0}{2i} \left ( \fourier{f}{1}(\bi{r},t) - \fourier{f}{-1}(\bi{r},t)  \right) \, .
 \end{eqnarray}
 \endnumparts
The second Fourier coefficients are related to the nematic order parameter $\abs{\mean{e^{2i\varphi}}}$ and the symmetric temperature tensor, as defined in \cite{romanczuk_mean-field_2012,grossmann_active_2012,romanczuk_swarming_2012}, which is a measure for the width of the velocity distribution (see \ref{sec:A:Temperature} for details). The dynamics of the temperature tensor is not derived explicitly in this context. 

In the following, the dynamics of the density and the momentum field is deduced from \eqnref{eqn:hierarchie:fourier_varphi:allg}. Therefore, the Fourier coefficients of the force $\fourier{Q}{n}(\bi{r},t)$ are approximatively calculated by expanding $p(\bi{r}+\bi{r}_{ji},\varphi,t)$ into a multidimensional 
Taylor series for small $\abs{\bi{r}_{ji}}$ \cite{aranson_tsimring_2005}, substituting \eqnref{eqn:angular:fourierseries:OPPDF} into \eqnref{eqn:self:consistent:force} and evaluating the remaining integral. Here, we consider all terms up to the second order of the Taylor expansion. This approximation holds for spatially slowly varying densities, i.e. large system size compared to the interaction radius $l_s$. 

Again, the Fokker-Planck equation turns into an infinite hierarchy of equations \eqnref{eqn:hierarchie:fourier_varphi:allg} in Fourier space. An appropriate closure scheme is used in order to consider only the first Fourier coefficients. Close to the order-disorder transition, the degree of polar order is assumed to be small, i.e. $\abs{\bi{w}}/(s_0 \rho) \propto \epsilon$, where $\epsilon$ is a small number and the particle density is locally close to the homogeneous distribution. Furthermore, the dynamics of the Fourier coefficients \eqnref{eqn:hierarchie:fourier_varphi:allg} suggests, that $\abs{\fourier{f}{1}}$ is larger than $\abs{\fourier{f}{n}}$ with $\abs{n}>1$, because of the damping term in \eqnref{eqn:hierarchie:fourier_varphi:allg} proportional to $n^2$, leading to the scaling relations
  \begin{eqnarray}
   \label{eqn:scaling:closure:scheme}
   \rho(\bi{r},t) - \rho_0 \propto \epsilon \, , \quad 
   \nabcp \propto \epsilon \, , \quad \partial_t \propto \epsilon \, , \quad \fourier{f}{n} \propto \epsilon^{\abs{n}} \, . 
  \end{eqnarray}
In \cite{peshkov_continuous_2012} it is argued, that the scaling ansatz of the temporal and spatial derivatives reflects the propagating nature of the system. This closure scheme described above was already used in \cite{bertin_boltzmann_2006,peshkov_continuous_2012,aranson_tsimring_2005,farell_marchetti_2012} and in \cite{peshkov_nonlinear_2012} in a modified way for nematic particles. A system of three nonlinear partial differential equations is obtained by keeping all terms up to the order $\epsilon^3$.
 \eqnlabel{eqn:gaskinetic:fourier:space}\numparts
  \begin{eqnarray}
			\fl	\frac{\partial \rho}{\partial t} &= - s_0 \left ( \nabcp \fourier{f}{1}^* + \nabcp^{*} \fourier{f}{1} \right) \label{eqn:continuity:eqn:fourier} \\
			\fl	\frac{\partial \fourier{f}{1}}{\partial t} &= \left (\xi_1 \rho - \xi_5 \right) \fourier{f}{1} - \xi_1 \fourier{f}{2} \fourier{f}{1}^{*} + \left ( \frac{32}{45} \xi_3 \rho - s_0 \right) \nabcp^* \fourier{f}{2} + \left ( \frac{32}{9} \xi_3 \rho - \xi_2 \rho - s_0 \right) \nabcp \rho \nonumber \\
			\fl	&\quad  - \frac{64}{45} \xi_3 \fourier{f}{1}^* \nabcp \fourier{f}{1} + \frac{\xi_4 \rho}{2} \left ( \Delta \fourier{f}{1} + 2 \nabcp^2 \fourier{f}{1}^* \right) \label{eqn:momentum:eqn:fourier}\\
			\fl \frac{\partial \fourier{f}{2}}{\partial t} &= 2 \left [ -2 \xi_5 \fourier{f}{2} + \xi_1 \fourier{f}{1}^2 - \left ( \frac{s_0}{2} + \frac{64}{45} \xi_3 \rho \right) \nabcp \fourier{f}{1} + \left ( \frac{32}{9} \xi_3 - \xi_2 \right) \fourier{f}{1} \nabcp \rho - \xi_4 \rho \nabcp^2 \rho \right ] \label{eqn:f2:eqn:fourier}
  \end{eqnarray}
 \endnumparts
The following coefficients are introduced:
\eqnlabel{eqn:xi} 
 \numparts
  \begin{eqnarray}
				\xi_1 =& \frac{\pi}{4} \int_{l_c}^{l_s} \rmd r_{ji} \, r_{ji} \left [ \mu_m(r_{ji}) - \mu_a(r_{ji}) \right ]	\, ,		\\
				\xi_2 =& \frac{\pi}{s_0} \int_{0}^{l_c} \rmd r_{ji} \, r_{ji}^2 \, \mu_r(r_{ji})	\, , \\
				\xi_3 =& \frac{1}{\pi} \int_{l_c}^{l_s} \rmd r_{ji} \, r_{ji}^2 \left [ \mu_m(r_{ji}) + \mu_a(r_{ji}) \right ] \, , \\
			 \xi_4 =& \frac{\pi}{8} \int_{l_c}^{l_s} \rmd r_{ji} \, r_{ji}^3 \left [ \mu_m(r_{ji}) - \mu_a(r_{ji}) \right ] \, , \\
 			\xi_5 =& \frac{D_{\varphi}}{s_0^2} \, .
  \end{eqnarray}
 \endnumparts
\Eref{eqn:continuity:eqn:fourier} is simply the continuity equation representing the conservation of the particle number $N$. 

\subsection{Hydrodynamic Limit}
\label{subsec:hydrodynamic_limit}

Another simplification of the system \eref{eqn:gaskinetic:fourier:space} is reached by adiabatically eliminating $\fourier{f}{2}$, i.e. $\partial_t \fourier{f}{2} \approx 0$, where
  \begin{eqnarray}
    \label{eqn:adiabat:elim:f2}
    \fl 			\fourier{f}{2} \approx \frac{1}{2 \xi_5} \left [ \xi_1 \fourier{f}{1}^2 - \left ( \frac{s_0}{2} + \frac{64}{45} \xi_3 \rho \right) \nabcp \fourier{f}{1} \right ]
  \end{eqnarray}
follows. This approximation is only valid if the first Fourier mode relaxes considerably slower than $\fourier{f}{2}$. The assumption of a time scale separation between the first and second Fourier mode can be only be justified if $\abs{\xi_1 \rho -\xi_5}\ll 4\xi_5$, which is equivalent to $-6\ll \kappa\ll10$. Furthermore $\xi_5 \propto D_{\varphi} > 0$ has to hold.

 Please note, that all terms of the order $\epsilon^3$ were dropped in \eref{eqn:adiabat:elim:f2} according to the scaling relations \eref{eqn:scaling:closure:scheme}. 

By identifying Fourier coefficients with physical quantities \eqnref{eqn:complex:numbers:real:quantities}, familiar hydrodynamic equations are obtained. 
 \eqnlabel{eqn:hydrodynamic:theory:direct}
 \numparts
  \begin{eqnarray}
				\fl	\frac{\partial \rho}{\partial t} &= - \nabla \cdot \bi{w} \\
    \fl \frac{\partial \bi{w}}{\partial t} &= \left ( \lambda_1 - \eta_1 \abs{\bi{w}}^2  \right)\bi{w} + \lambda_2 \nabla \rho + \lambda_4 \, \Delta \bi{w} + \lambda_5 \nabla (\nabla \cdot \bi{w}) + \eta_2 (\bi{w} \cdot \nabla) \, \bi{w} \nonumber \\
    \fl	& \quad + \eta_3 \left ( \frac{1}{2} \nabla \abs{\bi{w}}^2 - \bi{w} \, (\nabla \cdot \bi{w}) \right) \label{eqn:hydrodynamic:theory:flow:field}
  \end{eqnarray}
 \endnumparts
The transport coefficients $\{\lambda_i,\eta_i\}$ as functions of $\xi_i$ \eqnref{eqn:xi} read
\eqnlabel{eqn:transport:coefficients}
\numparts
 \begin{eqnarray}
		\lambda_1 &= \xi_1 \rho - \xi_5 , \\
  \lambda_2 &= \frac{s_0}{2} \left ( \frac{32}{9} \xi_3 \rho - \xi_2 \rho - s_0 \right) , \\
		\lambda_4 &= \frac{1}{4} \left  [ \frac{s_0 - \frac{32}{45} \xi_3 \rho}{2 \xi_5} \left ( \frac{s_0}{2} + \frac{64}{45} \xi_3 \rho \right) + \xi_4 \rho \right ]  ,\\
		\lambda_5 &= \frac{\xi_4 \rho}{2},  \\
		\eta_1 &= \frac{\xi_1^2}{2 \xi_5 s_0^2},  \\
		\eta_2 &= \frac{3 \xi_1}{8 s_0 \xi_5} \left ( \frac{256 \xi_3 \rho}{135} - s_0 \right) - \frac{32 \xi_3}{45 s_0},  \\
		\eta_3 &= \frac{5 \xi_1}{8 \xi_5} - \frac{32 \xi_3}{45 s_0}\ . 
\end{eqnarray}
\endnumparts
The hydrodynamic equations \eqnref{eqn:hydrodynamic:theory:direct} are comparable in structure to the theory of Toner and Tu \cite{toner_long-range_1995,toner_flocks_1998}.  

In the theory of Toner and Tu, the coefficients $\eta_2$ and $\eta_3$ are assumed to be nonzero phenomenological parameters due to the absence of Galilean invariance \cite{toner_flocks_1998}. Indeed, in our model, $\eta_2$ and $\eta_3$ can take both, negative and positive values. Likewise, $\lambda_2$ can be positive and negative: A positive $\lambda_2$-term leads to a flow into the direction of higher densities and consequently to clustering, as shown below. The damping coefficient $\eta_1$ is positive for all microscopic parameters, consistently. The sign of the different transport coefficients is discussed in \ref{sec:App:transport:coefficients} in detail. 

The stability of the hydrodynamic equations \eref{eqn:hydrodynamic:theory:direct} requires that $\lambda_4$ and $\lambda_5$ are greater than zero. In this region, the hydrodynamic equations describe collectively moving bands and the theory of Toner and Tu applies. In contrast to the original Vicsek model, in our model the coefficients $\lambda_4$ and $\lambda_5$ may change signs. As shown in \cite{wensink_meso-scale_2012}, a negative $\lambda_4$ may lead to an instability of spatially homogeneous states. In this case, higher order derivatives are needed in \eref{eqn:hydrodynamic:theory:flow:field} in order to guarantee the stability of the dynamics\footnote[6]{The derivation of those terms goes beyond the scope of the present work. }. Moreover, the time scale seperation of the first and second Fourier mode is only justified for $\xi_1 \rho \approx \xi_5$, i.e. $\lambda_1 \approx 0$. For the remaining analysis of the hydrodynamic equations, we will assume the stability and validity of our hydrodynamic theory \eref{eqn:hydrodynamic:theory:direct}.

Let us begin the analysis of \eref{eqn:hydrodynamic:theory:direct} by neglecting the spatial derivatives and analyzing the fixed points of 
  \begin{eqnarray}
    \frac{\rmd \bi{w}}{\rmd t} = \left ( \lambda_1 - \eta_1 \abs{\bi{w}}^2  \right)\bi{w} \, .
  \end{eqnarray}
That is the standard symmetry-breaking term present in all models of collective motion \cite{toner_flocks_1998,bertin_microscopic_2009,wensink_meso-scale_2012,ihle_kinetic_2011,aranson_tsimring_2005,farell_marchetti_2012,peshkov_continuous_2012}. Apparently, two spatially homogeneous solutions exist: First, $\bi{w}_0 = 0$ corresponding to a disordered phase and vanishing center of mass velocity. Second,
  \begin{eqnarray}
    \label{eqn:hydrodynamic:ordered:solution}
    \bi{w}_1 = \sqrt{\frac{\lambda_1}{\eta_1}} \, \bi{e} \, ,
  \end{eqnarray}
where $\bi{e}$ denotes an arbitrary unit vector reflecting the isotropy of the system. Without loss of generality we set ${\bi e}={\bi e}_x$. For $\lambda_1 < 0$, only the fixed point $\bi{w}_0$ exists and is stable against spatially homogeneous perturbations. For $\lambda_1 \ge 0$, $\bi{w}_1$ is a second fixed point corresponding to an ordered phase with nonzero mean-speed (swarming phase). 
\begin{figure}
\begin{center}
\centering\includegraphics[width=0.5\textwidth]{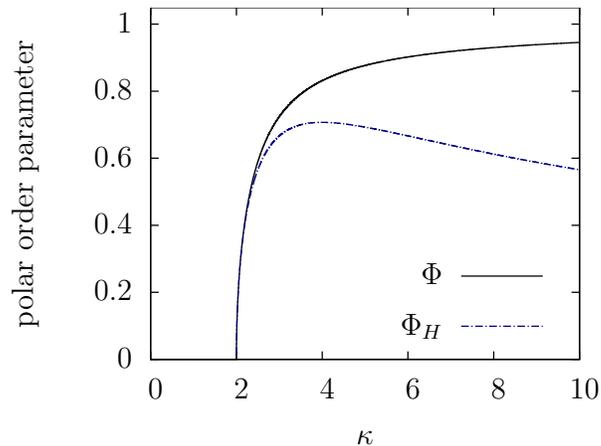}
\caption{Comparison of the order parameter for the spatially homogeneous obtained as an exact solution of the nonlinear Fokker-Planck equation from (\ref{eqn:bestimmungsgleichung:order:parameter}) and the hydrodynamic theory (\ref{eqn:polar:order:par:hydro}).
	\label{fig:comp_order}}
\end{center}
\end{figure}

The supercritical pitchfork bifurcation in $\lambda_1 = 0$, i.e. 
  \begin{eqnarray}
     \label{eqn:critical:line:order}
     \int_{l_c}^{l_s} \rmd r_{ji} \, r_{ji} \left( \mu_m(r_{ji}) - \mu_a(r_{ji}) \right) = \frac{4 D_{\varphi}}{\pi \rho_0 s_0^2} \, \mbox{, }
  \end{eqnarray}
corresponds to the order-disorder transition \eref{eqn:scaling:crit:point:FPE} described in section \ref{subsec:MFT}. By introducing the dimensionless coupling strength $\kappa$ \eref{eqn:dimless:kappa} again, 
 \begin{eqnarray}
   \label{eqn:dimless:kappa:hydro}
   \kappa = \frac{\pi \rho_0 s_0^2}{2 D_{\varphi}} \int_{l_c}^{l_s} \rmd r_{ji} \, r_{ji} \left ( \mu_m(r_{ji}) - \mu_a(r_{ji}) \right) \, , 
 \end{eqnarray}
the polar order parameter $\Phi_H = \abs{\bi{w}_1}/(s_0 \rho_0)$ as described by the hydrodynamic theory can be written as follows\footnote[7]{The density $\rho(\bi{r},t)$ was approximated by $\rho_0$ for that argument in accordance with \eref{eqn:scaling:closure:scheme}. }: 
 \begin{eqnarray}
   \label{eqn:polar:order:par:hydro} 
     \Phi_H = 2 \,  \frac{\sqrt{\kappa -2}}{\kappa} \, .
 \end{eqnarray}
Close to the critical point $\kappa_c = 2$, one recovers the correct scaling behaviour $\Phi_H \simeq \sqrt{\kappa -2}$ \eref{eqn:scaling:crit:point:FPE}. However, for large $\kappa$ the polar order parameter $\Phi_H$ tends to zero according to \eref{eqn:polar:order:par:hydro} as shown in figure \ref{fig:comp_order}. 
This is a consequence of the approximations made, namely the elimination of higher Fourier modes. If one allows a maximal relative error $(\Phi-\Phi_H)/\Phi$ of $5\%$, an upper bound for $\kappa$ is found numerically: $\kappa \le 2.617$. 

\Eref{eqn:critical:line:order} defines the transition line in the $(\mu_m$,$\mu_a)$-parameter space. Interestingly, only the integrated interaction strengths are important, the exact functional dependence on the distance between two particles $\mu_{m,a}=\mu_{m,a}(r_{ji})$ does not matter. Furthermore, collective motion is possible in both situations, pure attraction to particles moving away ($\mu_m > 0$, $\mu_a = 0$) and pure repulsion from approaching particles ($\mu_m = 0$, $\mu_a < 0$), respectively. Since the critical coupling parameter $\kappa_c=2$ is positive, polar order cannot be found in the anti-alignment regime. 

\subsection{Stability of the spatially homogeneous solutions}
We proceed with the analysis of homogeneous solutions  against spatial perturbations. At first, let us consider the stability of the disordered, homogeneous solution with $\bi{w}_0=0$. In order to investigate the stability of $\bi{w}_0$, the ansatz
 \numparts
  \begin{eqnarray}
					\rho(\bi{r},t)   &= \rho_0 + \delta \rho(\bi{r},t) \, , \\
		 		\bi{w}(\bi{r},t) &= \delta \bi{w}(\bi{r},t) \, 
  \end{eqnarray}
 \endnumparts
is inserted in \eref{eqn:hydrodynamic:theory:direct} and the resulting equations are linearized in the perturbations which are assumed to be small. The linearized equations read 
 \numparts
  \begin{eqnarray}
				\frac{\partial \delta \rho}{\partial t}   &= - \nabla \cdot \delta \bi{w} \, , \\
				\frac{\partial \delta \bi{w}}{\partial t}	&= \lambda_1 \delta \bi{w} + \lambda_2 \nabla \delta \rho + \lambda_4 \, \Delta \delta \bi{w} + \lambda_5 \nabla (\nabla \cdot \delta \bi{w})			\, .
  \end{eqnarray}
 \endnumparts
Inserting the exponential ansatz
 \eqnlabel{eqn:expon:ansatz:stability}
 \numparts
  \begin{eqnarray}
		 		\delta \rho(\bi{r},t)   &= \zeta_{\rho} \exp \left (\sigma t + i \bi{k}\cdot \bi{r} \right) \, , \\
		 		\delta \bi{w}(\bi{r},t) &= \boldsymbol{\zeta}_{\bi{w}} \exp \left (\sigma t + i \bi{k}\cdot \bi{r} \right)
  \end{eqnarray}
 \endnumparts
yields the dispersion relation $\sigma(\bi{k})$. Since $\bi{w}_0$ is isotropic, the eigenvalues $\sigma$ are independent of the direction of the wavevector. 
 \eqnlabel{eqn:hydrodynamic:eigenvalues:approx}
 \numparts
  \begin{eqnarray}
					\fl \sigma_1     &= \lambda_1 - \lambda_4 k^2 \, , \\
					\fl \sigma_{2,3} &= \frac{1}{2} \left [ \lambda_1 - \left ( \lambda_4 + \lambda_5 \right) k^2 \pm \sqrt{\left [\lambda_1 - (\lambda_4+\lambda_5) \, k^2 \right]^2 + 4 k^2 \lambda_2 } \, \right ] \, .  
  \end{eqnarray}
 \endnumparts
Provided all eigenvalues are less than zero, the disordered spatially homogeneous state is stable. The first eigenvalue $\sigma_1$ is equal to $\lambda_1$ for $k = 0$, i.e. the stability of $\bi{w}_0$ is determined by the sign of $\lambda_1$. That is the destabilization of the homogeneous state as discussed before. For $\lambda_4>0$, the first eigenvalue $\sigma_1$ is monotonically decreasing. However, the eigenvalue $\sigma_1$ is monotonically increasing for $\lambda_4 < 0$. In this case, the hydrodynamic equations \eref{eqn:hydrodynamic:theory:direct} loose their validity. This unphysical behaviour is due to the restriction to second derivatives when the integral \eref{eqn:self:consistent:force} was calculated. Furthermore, it is problematically that the Taylor coefficients have alternating signs so that Taylor polynomials converge slowly as argued at the end of section \ref{sec:stability:in:fourier}. To predict the behaviour of the system in that case, one needs to consider higher order terms. Unfortunately, the number of terms and Fourier coefficients that has to be considered to be consistent with the scaling ansatz \eref{eqn:scaling:closure:scheme}, grows rapidly. 

Nevertheless, it is possible to predict the stability of $\bi{w}_0$ by analyzing those eigenvalues, which tend to zero for small wavenumbers (hydrodynamic modes \cite{toner_flocks_1998}). For small wavenumbers, the eigenvalue $\sigma_2$ ($\sigma_2 > \sigma_3$) is expanded in a Taylor series:
  \begin{eqnarray}
    				\sigma_2 \simeq - \frac{\lambda_2}{\lambda_1} \, k^2 + \frac{\lambda_2^2 - \lambda_1 \lambda_2 \left ( \lambda_4 + \lambda_5 \right) }{\lambda_1^3} \, k^4 + \mathcal{O}(k^6) \, .
  \end{eqnarray}
Close to the order-disorder transition line, $\lambda_1$ is approximately zero. Therefore it is sufficient to keep the leading order terms in $1/\lambda_1$ \cite{bertin_microscopic_2009}:
  \begin{eqnarray}
        \label{eqn:ungeord:eigenval:hydro:small:k:1}
    				\sigma_2 \simeq - \frac{\lambda_2}{\lambda_1} \, k^2 + \frac{\lambda_2^2}{\lambda_1^3} \, k^4 + \mathcal{O}(k^6) \, .
  \end{eqnarray}
Suppose, $\bi{w}_0$ is stable against spatially homogeneous perturbations, i.e. $\lambda_1 < 0$. In that case, a long-wavelength instability emerges for positive $\lambda_2$. Considering \eref{eqn:hydrodynamic:theory:direct}, this instability is reasonable: for $\lambda_2 > 0$, inhomogeneities in the particle density will lead to a flow in the direction of the density gradients while for $\lambda_2 < 0$ density inhomogeneities decay due to inverse flows. The amplification of density gradients will lead to an agglomeration of particles (cf. figure \ref{fig:snapshots}E). This hypothesis is supported by numerical simulations of the microscopic model, as shown in the next section and in \cite{romanczuk_swarming_2012}. A similar instability due to the gradient term in \eref{eqn:hydrodynamic:theory:direct} is also known from other active matter systems \cite{farell_marchetti_2012}. Whereas in \cite{farell_marchetti_2012} the instability is due to the density-dependent motility of the particles, in our case the instability is referable to the selective interaction. The critical line is given by $\lambda_2 = 0$, i.e.
 \begin{eqnarray}
   \label{eqn:criterium:clustering}
   \fl \int_{l_c}^{l_s} \mbox{d} r_{ji}\, r_{ji}^2 \left ( \mu_m(r_{ji}) + \mu_a(r_{ji}) \right) = \frac{9 \pi}{16 s_0 \rho_0} \left ( \frac{s_0^2}{2} + \frac{\rho_0 \pi}{2} \int_0^{l_c} \rmd r_{ji} \, r_{ji}^2 \mu_r(r_{ji}) \right) \, .
 \end{eqnarray}

\begin{figure}
 \begin{center}
	\begin{minipage}{0.45\textwidth}
 	\begin{center}
  	\includegraphics[width=\textwidth]{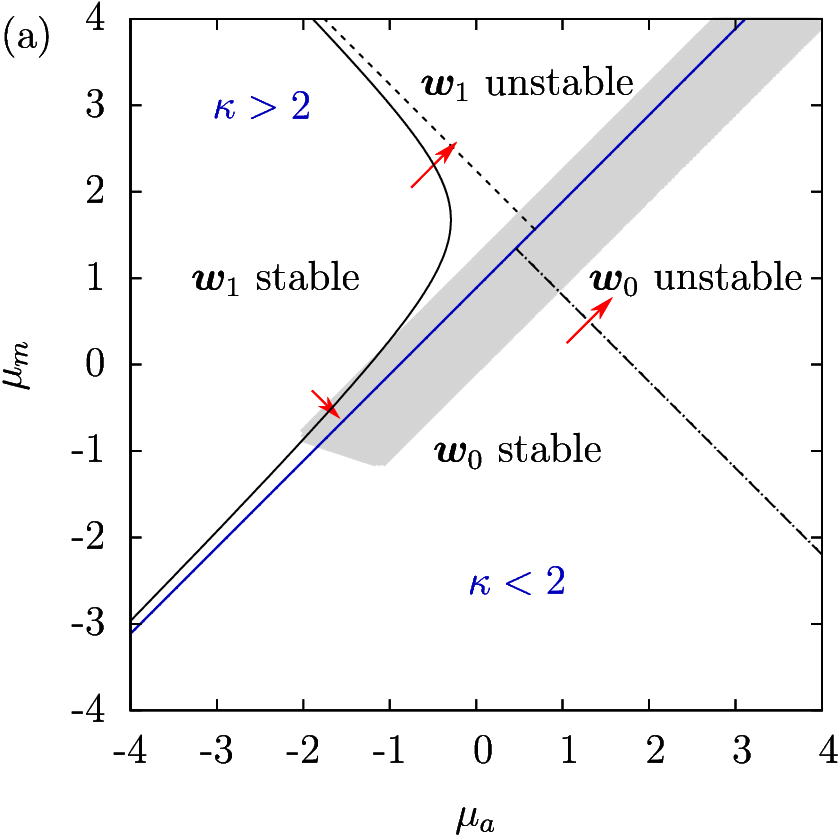}
 	\end{center}
	\end{minipage}
	\begin{minipage}{0.45\textwidth}
 	\begin{center}
  	\includegraphics[width=\textwidth]{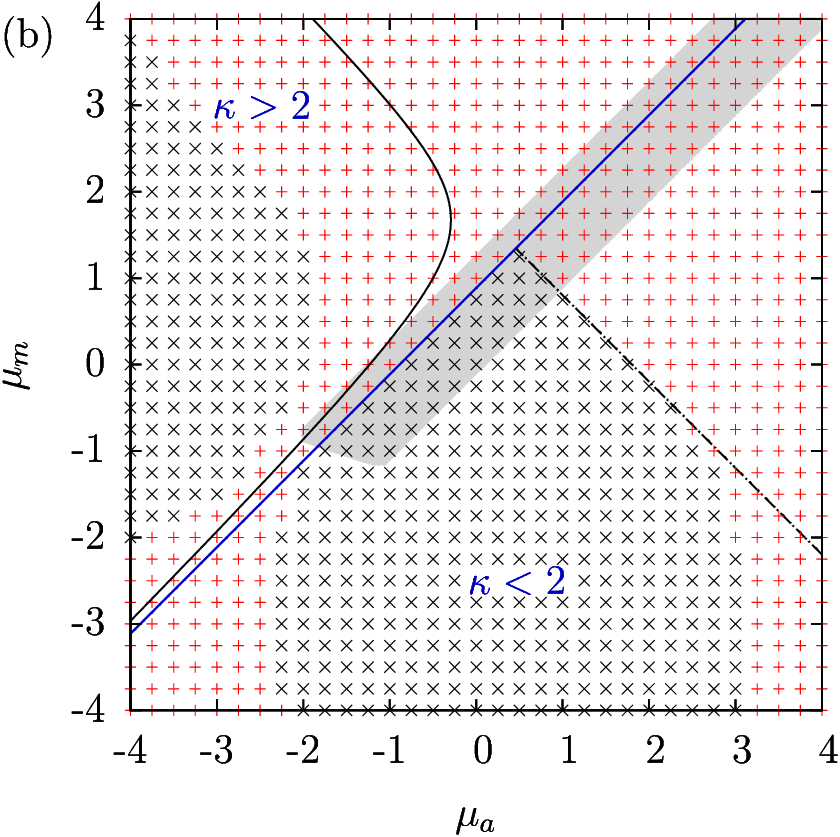}
 	\end{center}
	\end{minipage} 
  \caption{(a) Phase space as predicted by the hydrodynamic theory and (b) phase space in comparison to the predictions of the kinetic approach (see section \ref{sec:stability:in:fourier}). The arrows indicate long-wavelength instabilities in (a), black crosses indicate stable solutions, red plus signs indicate instabilities (see also figure \ref{fig:phase:space:stability:ana:kin:theo}) in figure (b). Below the critical blue (solid) line $\lambda_1 = 0$ (equal to $\kappa = 2$ \eref{eqn:critical:line:order}), only the solution $\bi{w}_0 = 0$ (among the spatially homogeneous solutions) exists. It is unstable above the critical (dotdashed) line $\lambda_2 = 0$ \eref{eqn:criterium:clustering}. Above the blue (solid) line, $\bi{w}_0$ is unstable. Besides, the solution $\bi{w}_1 = \sqrt{\lambda_1/\eta_1} \, \bi{e}$ exists. It is unstable beyond a hyperbola that is bounded by two asymptotes according to the hydrodynamic theory: black (dashed) line \eref{eqn:asymptotes:hyperbola:hydrdyn:geordnet:a} and blue (solid) line \eref{eqn:asymptotes:hyperbola:hydrdyn:geordnet:b}. The subspace of the parameter space where the hydrodynamic theory is valid is indicated by the grey-shaded region. Parameters: $D_{\varphi} = 0.5$; $\rho_0 = 1.5$; $l_c = 0.2$; $l_s = 1$; $s_0 = 1$; $\mu_r = 1$; $\mu_a$ and $\mu_m$ piecewise constant. }
  \label{fig:stability:hydrodyn}
 \end{center}
\end{figure}

The corresponding analysis of the stability of the spatially homogeneous ordered state for $|\bi{w}_1|>0$ using the full hydrodynamic equations is much more complicated. It is done in great detail for the original Vicsek model in \cite{bertin_microscopic_2009} for structurally identical hydrodynamic equations. In the following, the key points of the analysis are summarized. Moreover, an additional instability is found that is not present in the Vicsek model because in our model the transport coefficients $\lambda_2$, $\eta_2$ and $\eta_3$ may change their algebraic signs. Inserting the ansatz
 \numparts
  \begin{eqnarray}
					\rho(\bi{r},t)   &= \rho_0   + \delta \rho(\bi{r},t) \, , \\
		 		\bi{w}(\bi{r},t) &= \bi{w}_1 + \delta \bi{w}(\bi{r},t) \, 
  \end{eqnarray}
 \endnumparts
into \eref{eqn:hydrodynamic:theory:direct} and linearization yields 
 \numparts
  \begin{eqnarray}
				\fl \frac{\partial \delta \rho}{\partial t}   &= - \nabla \cdot \delta \bi{w} \, , \\
				\fl \frac{\partial \delta \bi{w}}{\partial t}	&= \left. \frac{\partial \lambda_1}{\partial \rho}\right |_{\rho_0} \delta \rho \, \bi{w}_1 - 2 \eta_1 \left (\bi{w}_1 \cdot \delta \bi{w} \right) \bi{w}_1 + \lambda_2 \nabla \delta \rho + \lambda_4 \Delta \delta \bi{w} + \lambda_5 \nabla (\nabla \cdot \delta \bi{w})		\nonumber \\
    \fl & \quad  + \eta_2 \left ( \bi{w}_1 \cdot \nabla \right) \delta \bi{w} + \eta_3 \left ( \nabla \left ( \bi{w}_1 \cdot \delta \bi{w} \right) - \bi{w}_1 \left ( \nabla \cdot \delta \bi{w} \right) \right) \, .
  \end{eqnarray}
 \endnumparts
Again, inserting \eqnref{eqn:expon:ansatz:stability} yields the growth rate $\sigma$ dependent on the wavevector $\bi{k}$. We restrict ourselves to longitudinal perturbations, meaning that the wavevector $\bi{k}$, the perturbation $\delta \bi{w}$ and the direction of collective motion $\bi{w}_0$ are pointing in the same direction, because we are interested in the emergence of collectively moving bands (cf. figure \ref{fig:snapshots}). The growth rate reads
 \begin{eqnarray}
   \sigma_{\pm} &= \frac{1}{2} \left [ -2 \lambda_1 - \left (\lambda_4 + \lambda_5 \right) k^2 + i \eta_2 \abs{\bi{w}_1} k \pm \sqrt{\alpha_1+i \alpha_2}  \, \right ] \, ,
 \end{eqnarray}
where the coefficients
\numparts
 \begin{eqnarray}
   \alpha_1 &= \left [ \left ( \lambda_4 + \lambda_5 \right) k^2 + 2 \lambda_1 \right]^2 - k^2 \left [ \left (\eta_2 \abs{\bi{w}_1} \right)^2 - 4 \lambda_2 \right] \, , \\
   \alpha_2 &= - 2 \abs{\bi{w}_1} k \left \{ 2 \left. \frac{\partial \lambda_1}{\partial \rho}\right |_{\rho_0} + \eta_2 \left [ \left ( \lambda_4 + \lambda_5 \right) k^2 + 2 \lambda_1 \right ] \right \}
 \end{eqnarray}
\endnumparts
were introduced. The real part of the largest eigenvalue reads as follows:
 \begin{eqnarray}
   \label{eqn:real:eigenval_hydrodyn:geordnet}
   \mbox{Re}(\sigma_{+})(k^2) &=  -\lambda_1 - \frac{\lambda_4 + \lambda_5}{2} k^2 + \sqrt{\frac{\alpha_1 + \sqrt{\alpha_1^2 + \alpha_2^2}}{8}} \, .
 \end{eqnarray}
Close to the order-disorder transition, $\lambda_1$ is small, with the result that it is sufficient to keep the lowest order in $1/\lambda_1$. The following expression is obtained for small wavenumbers \cite{bertin_microscopic_2009}: 
 \begin{eqnarray}
      \label{eqn:geord:eigenval:hydro:small:k:1}
     \mbox{Re}(\sigma_{+})(k^2) = \frac{\xi_1^2}{8\eta_1 \lambda_1^2} k^2 - \frac{5 \xi_1^4}{128 \eta_1^2 \lambda_1^5} k^4  + \mathcal{O}(k^6) \, .
 \end{eqnarray}
That implies, that for $\lambda_1 \gtrsim 0$, a long-wavelength instability emerges. By expanding \eref{eqn:real:eigenval_hydrodyn:geordnet} in a Taylor series for small $k$ (not assuming that $\lambda_1 \simeq 0$), one obtains
 \begin{eqnarray}
     \label{eqn:geord:eigenval:hydro:small:k:2}
     \mbox{Re}(\sigma_{+})(k^2) = \left ( \frac{1}{\eta_1} \left ( \frac{\xi_1}{\lambda_1} + \eta_2 \right) - \frac{\eta_2^2}{\eta_1} + \frac{4 \lambda_2}{\lambda_1} \right) \frac{k^2}{8} + \mathcal{O}(k^4) \, .
 \end{eqnarray}
In general, an instability appears, if the first Taylor coefficient becomes positive, because $\sigma_{+}(k \rightarrow 0) \rightarrow 0$. Unfortunately, it is not illuminating to derive the critical line $\mu_m = f(\mu_a)$, where the relevant Taylor coefficient is zero, analytically, because the resulting expressions are quite complicated. The numerical analysis reveals that the spatially homogeneous, ordered state is unstable beyond a hyperbola which is bounded by the two asymptotes (see also figure \ref{fig:stability:hydrodyn})
\eqnlabel{eqn:asymptotes:hyperbola:hydrdyn:geordnet}
\numparts
 \begin{eqnarray}
      \mu_m + \mu_a = \frac{135 \pi}{64 \rho_0 s_0 \left (l_s^3 - l_c^3 \right)} \left ( \frac{s_0^2}{2} + \frac{2 \pi \rho_0 l_c^3 \mu_r}{21} \right)  \label{eqn:asymptotes:hyperbola:hydrdyn:geordnet:a} \, ,\\
      \mu_m - \mu_a = \frac{8 D_{\varphi}}{\pi \rho_0 s_0^2 \left (l_s^2 - l_c^2 \right)} \label{eqn:asymptotes:hyperbola:hydrdyn:geordnet:b} \, . 
 \end{eqnarray}
\endnumparts
The instability of $\bi{w}_1$ for 
 \begin{eqnarray}
      \mu_m + \mu_a > \frac{135 \pi}{64 \rho_0 s_0 \left (l_s^3 - l_c^3 \right)} \left ( \frac{s_0^2}{2} + \frac{2 \pi \rho_0 l_c^3 \mu_r}{21} \right) 
 \end{eqnarray}
is due to the selective interaction, therefore it is not present in the original Vicsek model. The two asymptotes \eref{eqn:asymptotes:hyperbola:hydrdyn:geordnet} were derived using the approximation that the interaction strengths do not depend on the distance. The second line \eref{eqn:asymptotes:hyperbola:hydrdyn:geordnet:b} is equal to the critical line $\lambda_1 = 0$ \eref{eqn:critical:line:order}. 

As already mentioned, negative values of $\lambda_4$ and $\lambda_5$ may lead to additional instabilities, which go beyond the scope of this work, because the hydrodynamic equations are restricted to second derivatives.
However, neither \eref{eqn:ungeord:eigenval:hydro:small:k:1}, \eref{eqn:geord:eigenval:hydro:small:k:1} nor \eref{eqn:geord:eigenval:hydro:small:k:2} depend on $\lambda_4$ and $\lambda_5$. Hence, the spatially homogeneous states and instabilities described above are always present in our model.  

The predictions of the hydrodynamic theory, derived in this section, are compared to the predictions of the stability analysis using the kinetic description in section \ref{sec:stability:in:fourier} in figure \ref{fig:stability:hydrodyn}(b). The long-wavelength instability of $\bi{w}_0$, leading to clustering for $\lambda_2 > 0$ as well as the long-wavelength instability close to the critical point ($\kappa = 2$, $\lambda_1 = 0$) are confirmed by the kinetic approach. 

Within the range of validity of the hydrodynamic theory, no contradictions are found with the kinetic theory. However, figure \ref{fig:stability:hydrodyn}(b) indicates that the range of validity of the hydrodynamic theory does not cover all relevant parts of the parameter space. In particular, the destabilization of the ordered state $\bi{w}_1$ is not sufficiently well described by the hydrodynamic equations. Please note, that the hydrodynamic theory is only valid for $\lambda_4 > 0$ (guarantees the stability for large wavenumbers) and $0 \le \kappa \le 2.617$, where the lower bound is equivalent to $\lambda_5 > 0$ and the upper bound ensures small deviations $\Phi-\Phi_H$. Due to the last limitation, it is impossible to study the dynamics of the system for high order parameters, i.e. far away from the critical point. 

However, one can show numerically that the dispersion relations of the hydrodynamic theory \eref{eqn:hydrodynamic:theory:direct} and the full system \eref{eqn:hierarchie:fourier_complete:allg} coincide for small wavenumbers in the parameter range, where the hydrodynamic equations are valid. The hydrodynamic theory yields only three eigenvalues, which coincide with three eigenvalues given by the kinetic theory in the limit $k \rightarrow 0$. For large $k$, deviations occur due to the restriction to second derivatives (cf. scaling ansatz \eref{eqn:scaling:closure:scheme}). 

Nevertheless, the hydrodynamic theory yields important insights on the macroscopic behaviour of the system. The symmetry breaking, i.e. the existence of a collective motion mode is predicted. Moreover, the stability of the homogeneous solutions to the hydrodynamic equations can be analyzed analytically and it allows to obtain analytical results on the order-disorder transition line as well as the occurrence of clustering. Thus, the structure of the phase space is roughly estimated by the hydrodynamic theory. New instabilities are found due to changes in the algebraic signs of the coefficients $\lambda_2$ and $\eta_2$. Furthermore, the structure of the theory suggests, that all predictions of the theory by Toner and Tu \cite{toner_flocks_1998} are expected to arise in our system, such as giant number fluctuations. Within the range of validity of the hydrodynamic theory, the predictions of the kinetic theory, which is based on the mean-field assumption only (see section \ref{sec:kinetic} for details), are in agreement with the hydrodynamic theory, whose range of validity is known quantitatively. 

\subsection{Large-System Limit}
The hydrodynamic theory is simplified further, if the limit of large systems, i.e. large particle densities ($N\rightarrow \infty$) and small interaction regions ($l_s \rightarrow 0$) is considered, such that the number of particle within the interaction area is kept fixed: $\rho \, l_s^2 \approx \mbox{const}$. This approximation corresponds to a zeroth order Taylor expansion of the force \eref{eqn:self:consistent:force}: $p(\bi{r}+\bi{r}_{ji},\varphi_j,t) \approx p(\bi{r},\varphi_j,t)$. This approximation of ``ultra-local'' coupling is used in \cite{bertin_microscopic_2009,peshkov_continuous_2012} within a Boltzmann approach, where the collision integral is an integral over the particles orientation but not over relative distances. In that case, the hydrodynamic equations read
 \numparts
  \begin{eqnarray}
				\fl	\frac{\partial \rho}{\partial t} &= - \nabla \cdot \bi{w} \, ,\\
    \fl \frac{\partial \bi{w}}{\partial t} &+ \frac{3 s_0^2 \, \xi_1 }{8 D_{\varphi}} \, (\bi{w} \cdot \nabla) \, \bi{w} = \left [ \left ( \xi_1 \rho - \frac{D_{\varphi}}{s_0^2} \right) - \frac{\xi_1^2}{2 D_{\varphi}} \abs{\bi{w}}^2  \right]\bi{w} - \frac{s_0^2}{2} \nabla \rho + \frac{s_0^4}{16 D_{\varphi}} \, \Delta \bi{w} \nonumber \\
    \fl	& + \frac{5 s_0^2 \, \xi_1}{8 D_{\varphi}} \left ( \frac{1}{2} \nabla \abs{\bi{w}}^2 - \bi{w} \, (\nabla \cdot \bi{w}) \right) \, .
  \end{eqnarray}
 \endnumparts
Since $D_{\varphi}$ is always positive, the dynamics is stable for all microscopic parameters. The coefficient $\xi_1$ in front of the hydrodynamic term $(\bi{w} \cdot \nabla) \, \bi{w}$, which corresponds to the convective derivative in the equilibrium case, is positive for $\mu_m > \mu_a$ and is negative for $\mu_m<\mu_a$. The same holds true for the other ``hydrodynamic terms'' $\nabla \abs{\bi{w}}^2$ and $\bi{w} \, (\nabla \cdot \bi{w})$. Microscopically, the transition from positive $\xi_1$ to negative $\xi_1$ means, that the selective interaction turns from an effective alignment to an effective anti-aligning interaction, meaning that a focal particle aligns its velocity in an anti-parallel fashion with respect to the local mean-velocity (see figure \ref{fig:scheme}). Similar conclusion could possibly be drawn from \cite{farell_marchetti_2012,peruani_mean-field_2008}, if one allows the alignment strength to be negative. In that part of the parameter space, the nematic structures emerge in our model according to the individual based simulations discussed in the next section. 

\section{Comparison with Numerical Simulations}

\label{sec:simulations}

In order to analyze the stability of the disordered, homogeneous state, we have performed systematic numerical simulations of the individual-based model \eref{eq:eom_model}. The degree of collective motion was measured using the time averaged polar order parameter 
\begin{equation}\label{eqn:polar_op}
\mean{\Phi}_t = \mean{ \abs{ \frac{1}{N s_0} \sum_{i=1}^N {\bi v}_i(t) }}_{\! t} \, .
\end{equation}
Here $\langle\cdot\rangle_t$ represents a temporal average.
$\mean{\Phi}_t=1$ corresponds to perfect orientational order with all agents moving in the same direction, whereas a vanishing $\mean{\Phi}_t$ corresponds to a completely disordered system. 
Please note that $\mean{\Phi}_t=0$ can only be observed in the thermodynamic limit ($N\to\infty$). In a finite, disordered system we will measure a small, but finite $\mean{\Phi}_t\gtrapprox 0$ due to finite size fluctuations of the order $1/\sqrt{N}$. 

In order to measure the deviations from a spatially homogeneous state, we have subdivided the simulation domain into square cells of size $l_s \times l_s$ set by the sensory range. We used this spatial subdivision to calculate the spatial entropy function
\begin{eqnarray}
S=-\sum_{j, n_j \neq 0} p_j \log p_j = -\sum_{j, n_j \neq 0} \frac{n_j}{N}\log\frac{n_j}{N},
\end{eqnarray}
where the summation occurs over all occupied cells of the grid with $n_j\geq 1$ ($n_j$ is the number of particles in the $j$ths cell). This allows us to define the following spatial order parameter
\begin{equation}\label{eqn:spatial_op}
\mean{\Psi}_t = \mean{1-\frac{S}{S_{max}}}_t
\end{equation}
with $S_{max}$ being the maximal value of the spatial entropy corresponding to a homogeneous distribution of particles. $\mean{\Psi}_t=0$ corresponds to a perfectly disordered state, whereas $\mean{\Psi}_t>0$ indicates a spatially inhomogeneous distribution of particles (clusters, bands), where $\mean{\Psi}_t=1$ corresponds to the extreme situation where all particles are located in a single cell.

\begin{figure}
 \begin{center}
	$\rho_0=10.0$ \\
	\begin{minipage}{0.45\textwidth}
 	\begin{center}
	$\langle \Phi \rangle_t$ \\
  	\includegraphics[width=\textwidth]{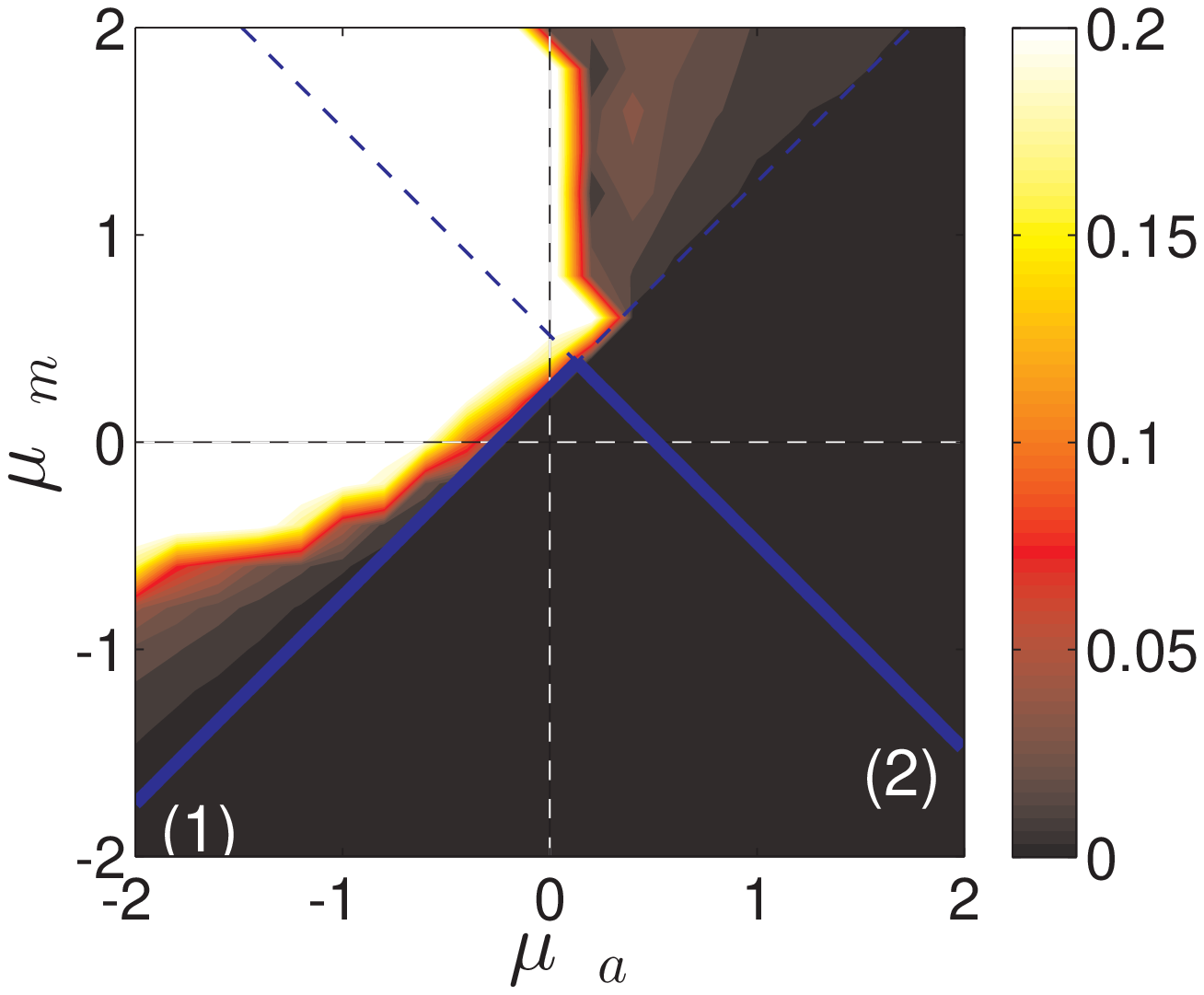}
 	\end{center}
	\end{minipage}
	\begin{minipage}{0.45\textwidth}
 	\begin{center}
	$\langle \Psi \rangle_t$ \\
  	\includegraphics[width=\textwidth]{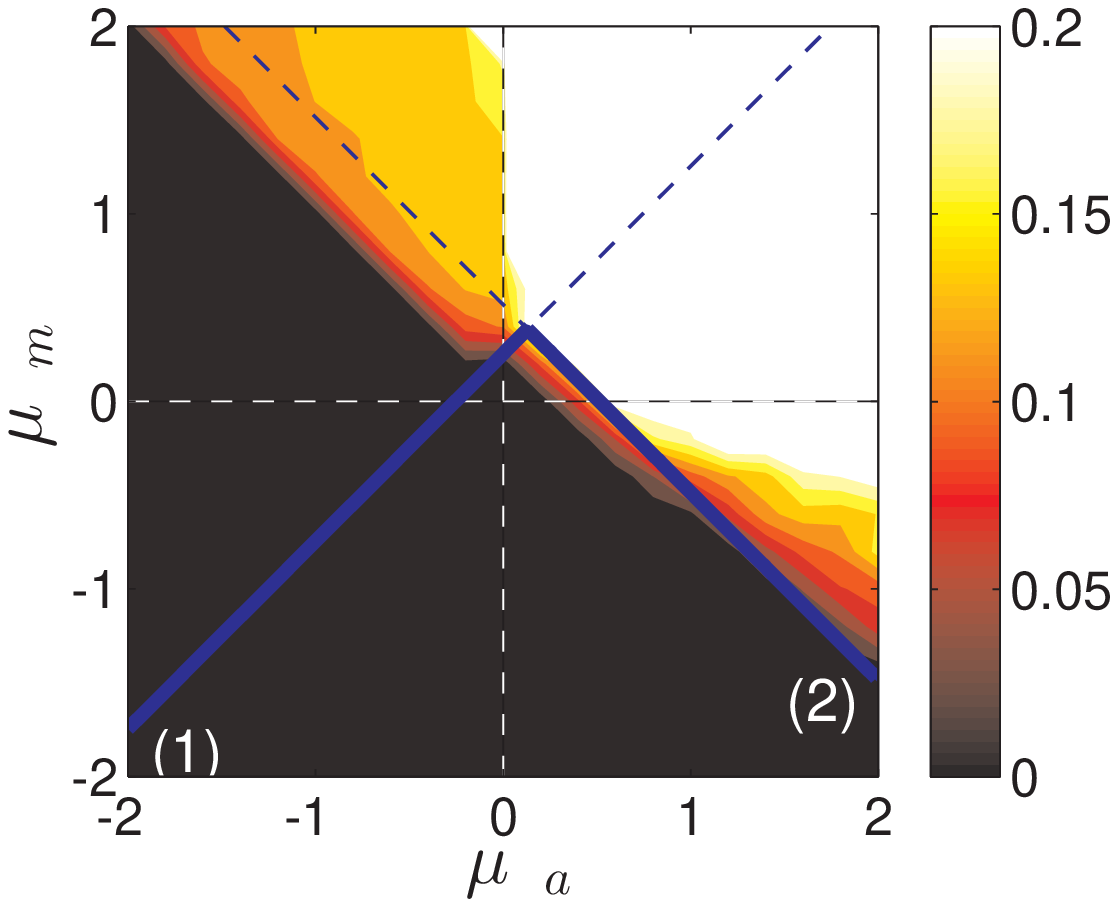}
 	\end{center}
	\end{minipage} 
	$\rho_0=2.5$\\
	\begin{minipage}{0.45\textwidth}
 	\begin{center}
	$\langle \Phi \rangle_t$ \\
  	\includegraphics[width=\textwidth]{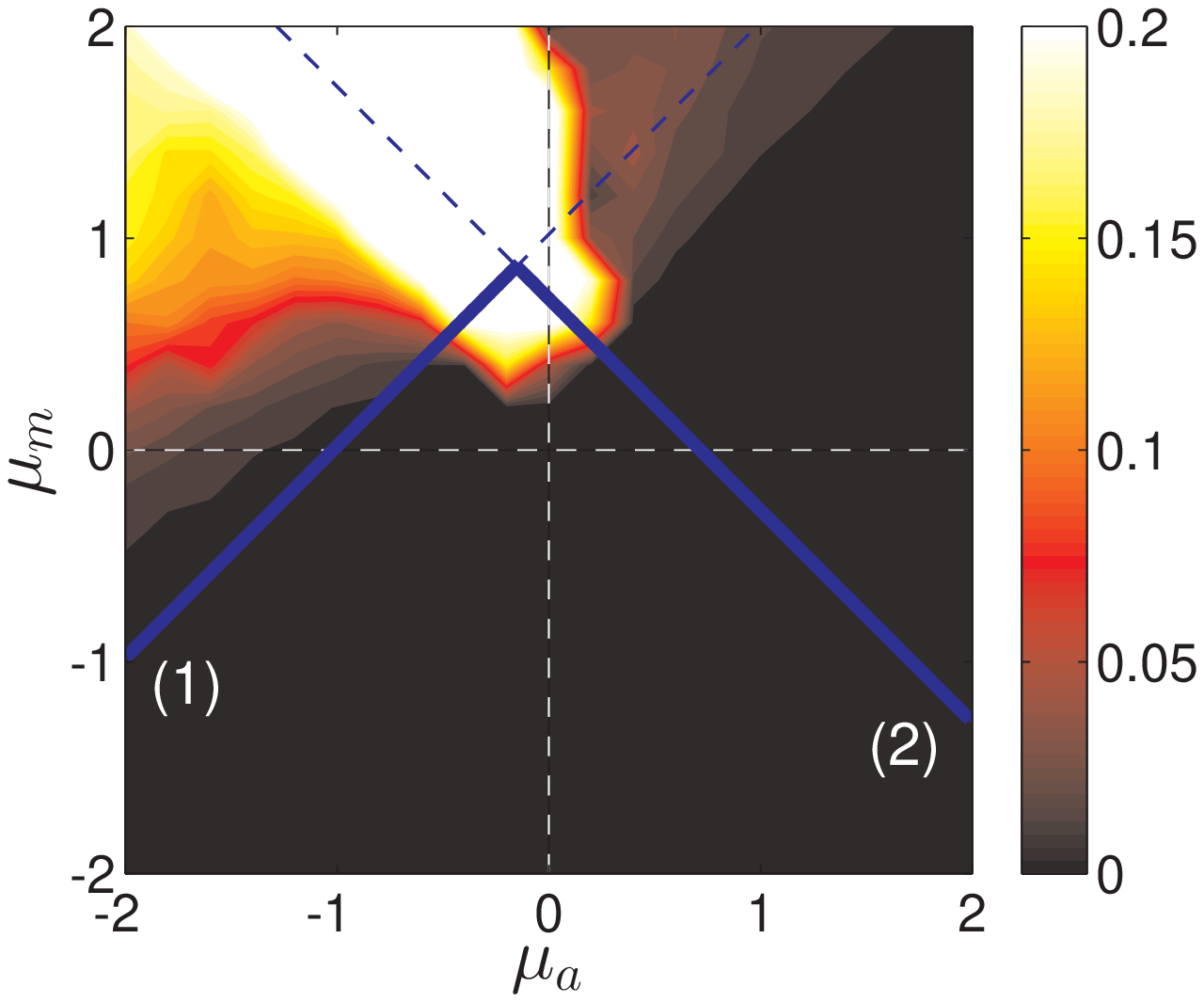}
 	\end{center}
	\end{minipage}
	\begin{minipage}{0.45\textwidth}
 	\begin{center}
	$\langle \Psi \rangle_t$ \\
  	\includegraphics[width=\textwidth]{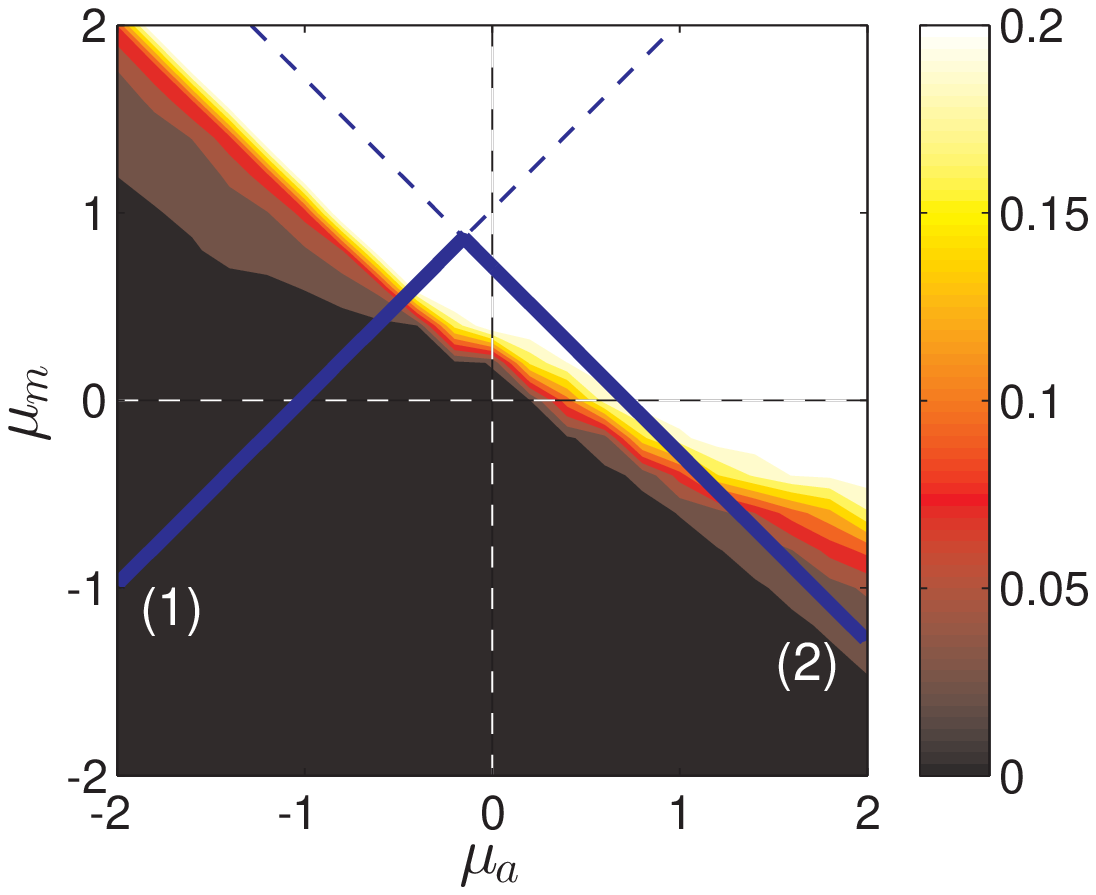}
 	\end{center}
	\end{minipage}
  \caption{{Comparison of numerical simulations of the individual-based model with the predictions of the hydrodynamic theory on the stability of the disordered solution in the $(\mu_m,\mu_a)$-plane for different densities: $\rho_0=10.0$ ($L=20$, top) and $\rho_0=2.5$ ($L=40$, bottom). {\bf Left:} Average orientational order parameter $\langle\Phi\rangle_t$; {\bf Right:} 
Average spatial inhomogeneity order parameter $\langle\Psi\rangle_t$. 
 The solid blue (gray) lines show the critical lines for the instability of the disordered solution. Line (1) corresponding to the orientation instability \eref{eqn:critical:line:order}, whereas line (2) corresponds to the clustering instability \eref{eqn:criterium:clustering}.
 Other parameters: $D_\varphi=0.06$, $N = 4000$, $\mu_r = 5$, $l_c = 0.2$, $l_s = 1$, $s_0 = 0.25$, $dt=0.01$.} }
  \label{fig:simulation1}
 \end{center}
\end{figure}

\begin{figure}
 \begin{center}
	\begin{minipage}{0.49\textwidth}
 	\begin{center}
	$\langle \Phi \rangle_t$ \\
  	\includegraphics[width=\textwidth]{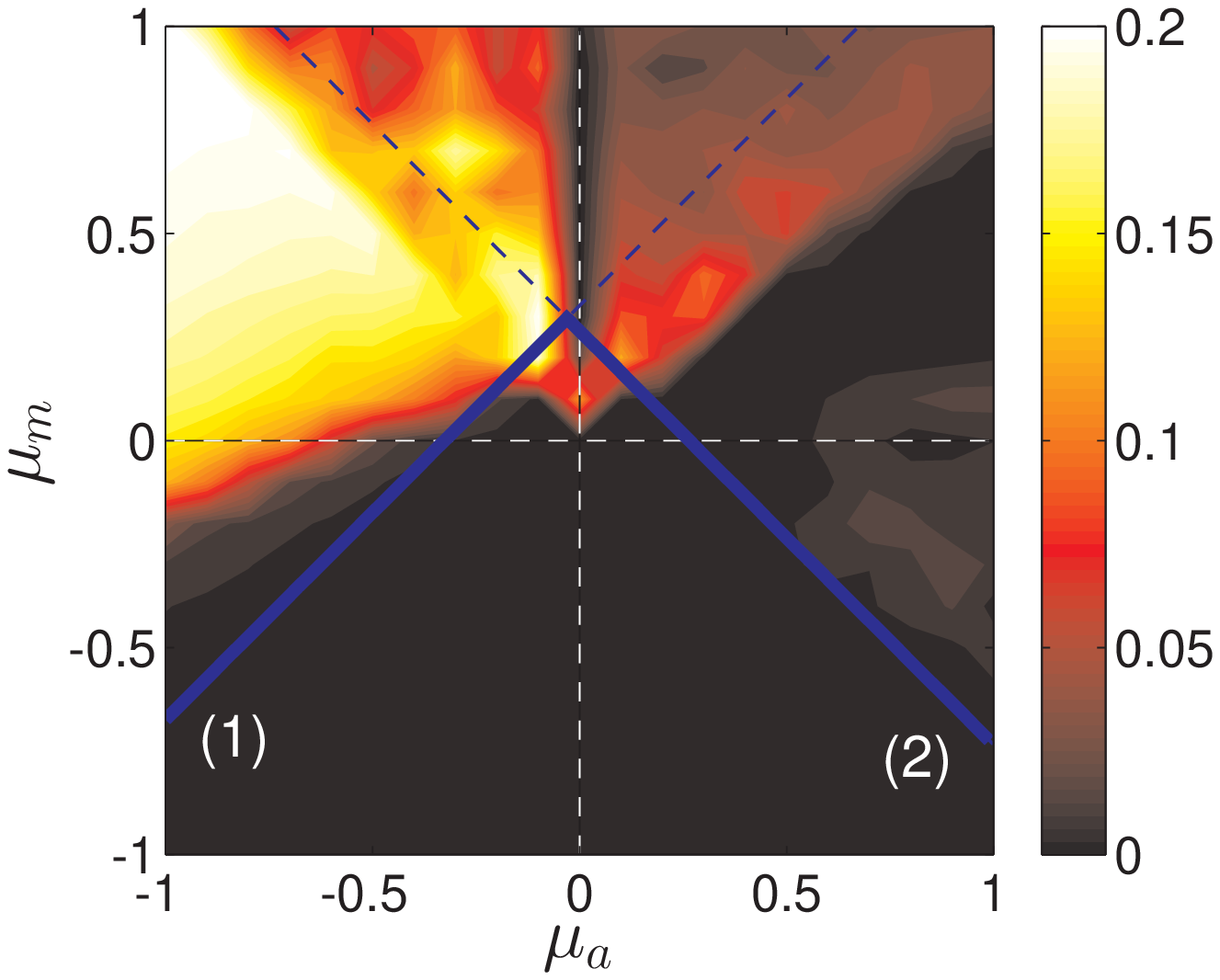}
 	\end{center}
	\end{minipage}
	\begin{minipage}{0.49\textwidth}
 	\begin{center}
	$\langle \Psi \rangle_t$ \\
  	\includegraphics[width=\textwidth]{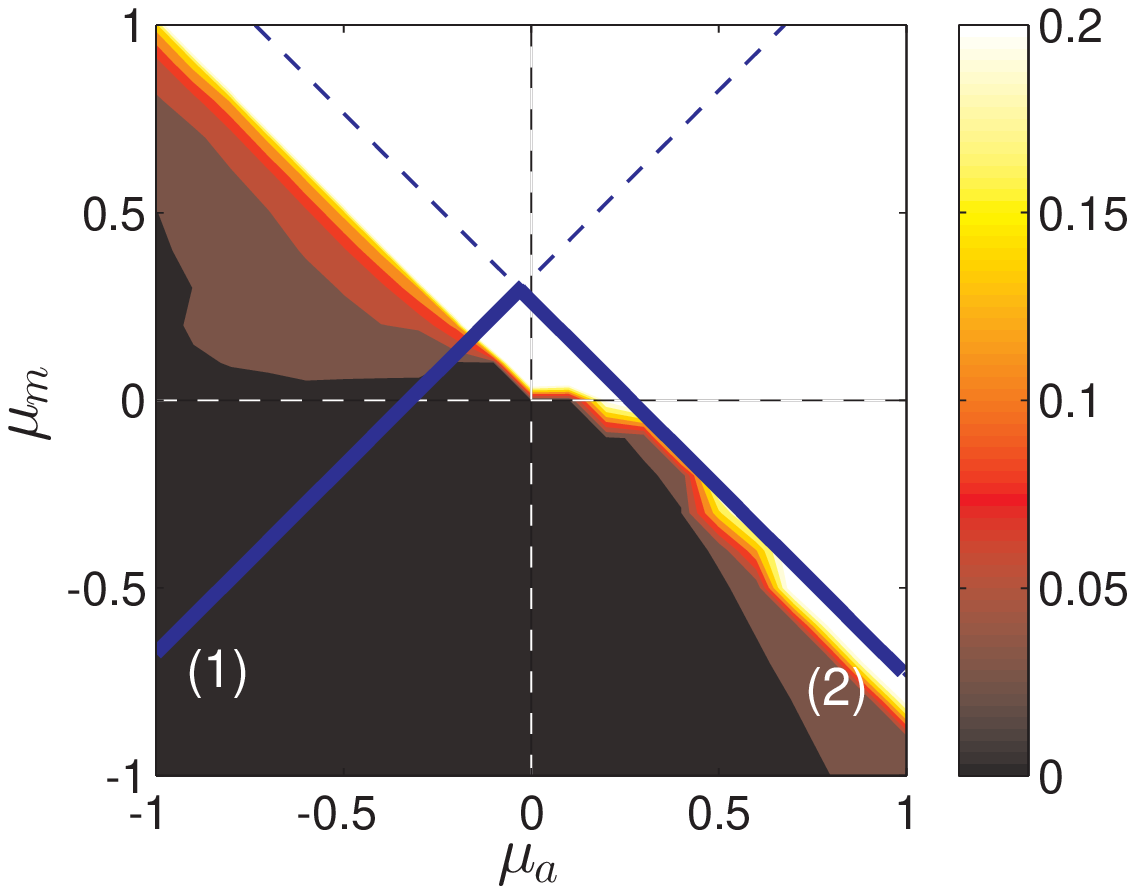}
 	\end{center}
	\end{minipage} 
\end{center}
\caption{Results of numerical simulations on the stability of the disordered solution in the $(\mu_m,\mu_a)$-plane for vanishing short-ranged repulsion ($l_c=0$). The spatial order parameter $\mean{\Psi}_t$ shows emergence of structures below the critical line (2). Other parameters: $N=4000$, $l_s=1$, $s_0=0.25$, $D_\varphi=0.02$, $L = 40$. \label{fig:vanishing_lc}}
\end{figure}

In order to test the stability of the disordered state, we have averaged the two order parameters over a time interval $\Delta t=1000$ after an initial time $t_{ini}=1000$. Please note, that for certain parameter values it is possible that the system has not reached a stationary state after $t=1000$. However, this initial time is sufficient to account for deviations from the homogeneous disordered state, which we are interested in. Accordingly, we are not interested in the actual stationary values of $\mean{\Phi}_t$ and $\mean{\Psi}_t$. Thus, we set the upper limit of the colour bar for both order parameters in figures \ref{fig:simulation1} and \ref{fig:vanishing_lc} to $0.2$. In consequence, all regions of parameter space, where $\mean{\Phi}_t \geq 0.2$ or $\mean{\Psi}_t\geq 0.2$ will appear white in figures \ref{fig:simulation1} \& \ref{fig:vanishing_lc}. 

The stochastic differential equations of the microscopic model were integrated with periodic boundary conditions using the stochastic version of the Euler algorithm with a numerical time step $dt=0.01$. This time steps is two orders of magnitude smaller than the average time scale of turning of individuals due to binary interactions for the range of interaction strengths in the simulations. Additional sample runs with smaller time steps did not show any significant differences in the simulation results. The initial condition for all simulations was the disordered, spatially homogeneous state.

For simplicity, we consider only constant interaction strengths (independent of the distance): $\mu_{m,a}(r_{ji})=\mu_{m,a}=\mbox{const.}$ 
Here, we focus on the analysis of the stability of the disordered, homogeneous solution with respect to variations of the interaction parameters $\mu_a$ and $\mu_m$. The analytical predictions of the hydrodynamic theory for the onset of instability of the disordered solution are represented by two intersecting critical lines in the $(\mu_a,\mu_m)$-plane perpendicular to each other (figure \ref{fig:simulation1}). The first one ($\mu_a \sim \mu_m$) corresponds to the orientational instability and onset of collective motion, cf. \eref{eqn:critical:line:order} and line (1) in figure \ref{fig:simulation1}, whereas the second one ($\mu_a \sim -\mu_m$) corresponds to the density instability associated with structure formation due to effective attraction between particles, cf. \eref{eqn:criterium:clustering} and line (2) in figure \ref{fig:simulation1}. The homogeneous, disordered solution is predicted to be linearly stable only ``below'' both critical lines.

The emergence of collective motion as well as the destabilization of the homogeneous density distribution is mostly confirmed by the numerical individual-based simulations. In specific parameter regions, deviations from the hydrodynamic theory are observed. 

In particular, we observe a high degree of collective motion in the effective-alignment region and strong deviations from the homogeneous spatial distribution of particles in the pure attraction regime as well as in the effective-alignment regime (in particular in the regime $\left (\mu_m,\mu_a \right) \in \left \{ \left (\mu_m,\mu_a \right) : |\mu_m|>|\mu_a| , \mu_m>0 \right \}$). This also confirms our previous results of comprehensive numerical investigations of related individual-based models \cite{romanczuk_collective_2009,romanczuk_swarming_2012}. 

Disagreements between simulations and the theoretical prediction (non-vanishing orientational order and/or clustering in simulations below the two critical lines) appear predominantly close to the intersection of the two critical lines and are stronger at low densities. They might be associated with the mean-field assumptions used in order to derive the hydrodynamic theory. On the one hand, at low densities the assumption of a continuous density of neighbours is strongly violated. On the other hand, correlations between particles may play an important role in the respective parameter range (likewise for high densities). Thus, the factorization of the N-particle PDF into a product of one-particle PDFs \eref{eqn:mean-field_assumpt} leads to a questionable approximation in that case. 

Another possible explanation for disagreement between theory and simulation is a breakdown of the homogeneous, disordered solution due to finite amplitude instabilities at parameters where this solution is still linearly stable.

\begin{figure}
 \begin{center}
 	\begin{center}
  	\includegraphics[width=0.9\textwidth]{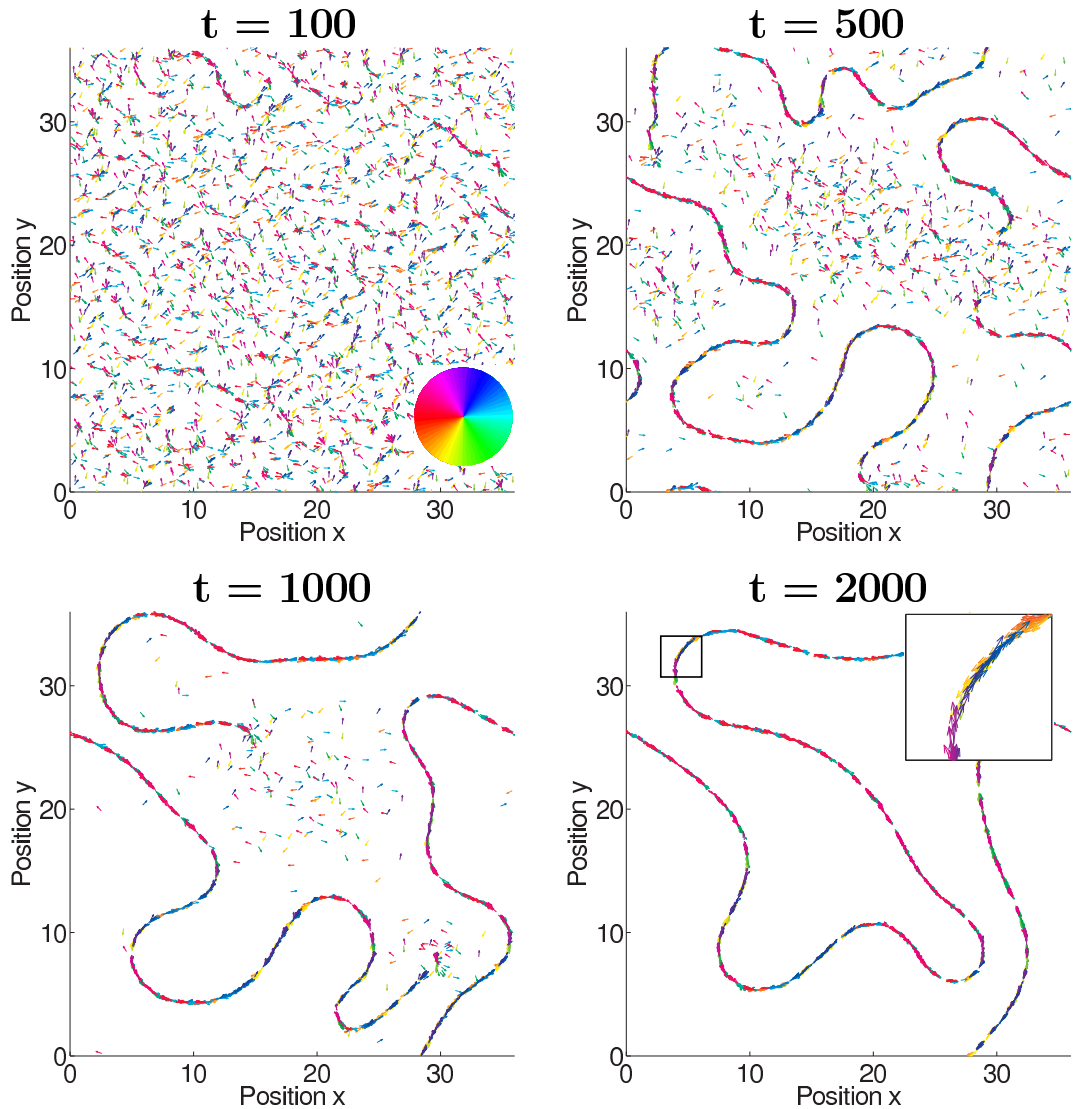}
 	\end{center}
\end{center}
\caption{A sequence of snapshots of one simulation for different times visualizing the formation of nematic filaments for $\mu_a=0.6$, $\mu_m=-0.4$. The inset in (D) shows a close up of the spatial region indicated by the square in the upper left corner. Other parameters: $N=4000$, $l_s=1$, $l_c=0$, $s_0=0.25$, $D_\varphi=0.02$, $L = 36$. \label{fig:snapshots_nematic}}
\end{figure}

For weak (or vanishing) short-ranged repulsion ($l_s\gg l_{c}$ or $\mu_r\ll|\mu_{m,a}|$), we observe inhomogeneous states without polar order far in the effective anti-alignment regime ($\mu_m<0$, $\mu_a>0$), clearly below the critical line predicted by the hydrodynamic theory as shown in figure \ref{fig:vanishing_lc}. This instability was missed in the previous study of the model  \cite{romanczuk_swarming_2012}.  
We were able to confirm it, using our kinetic approach (see section \ref{sec:kinetic}) by positive eigenvalues of the matrix determining the stability of the disordered solution at the respective parameter values. A close inspection of the dynamics of the individual-based model in this parameter region reveals the emergence of dense nematic filaments with particles moving in an anti-parallel fashion within the filament, whereby approximately half of the particles moves in either direction along the filament (see figure \ref{fig:snapshots_nematic}). 

The hydrodynamic theory derived in section \ref{sec:hydrodynamic} is not able to account for this kind of structures as it considers only the (polar) momentum field and not the nematic director field as a coarse-grained variable. An extension of the hydrodynamic theory by including an equation for the nematic director field may allow for identifying instabilities due to the onset of nematic order, however it goes beyond the scope of this work. We refer the reader to some recent theoretical works on active nematics \cite{ramaswamy_active_2003,baskaran_hydrodynamics_2008,ginelli_large-scale_2010,mishra_dynamic_2010,peshkov_nonlinear_2012}.

	\section{Discussion}


The self-propelled particle model with selective attraction-repulsion interaction shows a rich dynamical behavior. In the ``effective alignment'' regime, characterized by repulsion from approaching agents and attraction to moving away agents, the model behavior is closely related to the well-known Vicsek model: For repulsion from approaching particles equal to, or stronger than, the attraction to particles moving away, we typically observe the homogeneous flocking phase with giant number fluctuations predicted by Toner and Tu. However, in the opposite case, if the attraction to particles moving away starts to dominate, we observe strong clustering and effective phase separation, which is not observed in minimal models with constant speed and velocity alignment. 
Furthermore, by continuously varying model parameters, we can observe other phases such as unpolarized clusters for pure attraction or dense filamentous nematic structures in the effective anti-alignment regime.

In order to obtain a fundamental understanding of the phase diagram of the model, we derived first a kinetic description of the system based on the Fourier transform of the probability density function. The corresponding system of equations for the successive Fourier modes can be used for efficient numerical analysis of the linear stability of solutions of the nonlinear Fokker-Planck equation. We have analyzed the stability of the spatially homogeneous solutions. In addition, we have shown that the integration over the social forces yields Bessel functions of the first kind, which enter the matrix elements of the corresponding linearized system of differential equations in Fourier space. Due to the alternating Taylor coefficients of the corresponding Bessel functions, a closure approximation, corresponding to a finite order expansion, immediately leads to unphysical divergences at large wavenumbers $k$. 

Furthermore, we have derived a hydrodynamic theory by truncating the ``small-wavenumber-expansion'' at the second order. The resulting hydrodynamic equations are in agreement with the generic Toner and Tu theory of active matter. Our work establishes a direct link between the microscopic parameters of the individual-based model and the macroscopic parameters governing the behaviour of the coarse-grained hydrodynamic variables (density and momentum fields).  
Interestingly, the hydrodynamic parameters,  as for example $\eta_3$, which relates to a splay elasticity in corresponding equilibrium systems, or $\lambda_4$ and $\lambda_5$ which govern the relaxation of splay and bend fluctuations \cite{mishra_fluctuations_2010}, may change their sign in dependence on microscopic model parameters. Using the hydrodynamic theory, we can track down such sign changes to corresponding microscopic dynamics. Considering a limit of large system sizes, we reveal the importance of the change of sign of $\sigma_1$, which indicates the transition from an overall aligning effect of the social interaction to the opposite case where the interaction tend to anti-align interacting particles.  
Finally, a comparison between the eigenvalues obtained from hydrodynamic theory and the kinetic description, which takes higher orders into account, allows us to assess the validity of the hydrodynamic theory. 

We performed extensive simulations of the microscopic model based on stochastic differential equations focusing on the $(\mu_m,\mu_a)$-plane in the parameter space. The critical lines obtained from the hydrodynamic theory, where the disordered, spatially homogeneous solution becomes unstable, show good agreement with the numerical results at sufficiently high densities. However, at certain parameters clear deviations appear, as for example in the vicinity of the crossing of the two critical lines at low densities. 

This leads us to the important question of the validity of the approximations made during the coarse-graining. One common assumption is that of molecular chaos, which allows to factorize the N-particle PDF into a product of N one-particle PDFs. In models with collision-like interactions, the approximation is supposed to work best at low densities, where the mean-free path of particles between interactions is large\cite{bertin_boltzmann_2006,ihle_kinetic_2011}. However, in our case, we observe large deviations at low densities. This contradicting effect may be due to an approximation required to evaluate the integrals over the social forces. It relies on a Taylor expansion of the one-particle density function around the position of a focal individual. However at low densities, the interaction range $l_s$ is not the suitable coarse-graining scale as assumed implicitly in the formulation of the one-particle Fokker-Planck equation. Similar arguments have been put forward also in \cite{farell_marchetti_2012}. The impact of individual (angular) noise on the mean-field assumption may also be not straight forward. Intuitively, one would argue that uncorrelated individual noise terms always decrease correlations between interacting particles. However, for self-propelled particles, angular noise leads to a stronger localization of particles \cite{mikhailov_self_motion_1997,romanczuk_active_2012}, thus in principle also to a prolonged interaction between neighbours, which may in principle enhance multi-particle correlations.  

Furthermore, the systematic comparison of the prediction of the kinetic theory with the predictions of the hydrodynamic theory and numerical simulations revealed an unexpected additional instability. It corresponds to the emergence of dense filamentous structures with nematic order. Onset of nematic order is linked to the dynamics of the second Fourier amplitude, which was adiabatically eliminated in order to derive the hydrodynamic theory. Thus, the presented hydrodynamic theory cannot account for this instability, but it can be well traced by numerical evaluation of the linearized kinetic equations derived in section \ref{sec:kinetic}. 

So far, hydrodynamic equations of active matter were derived directly from minimal microscopic models of self-propelled particles with velocity-alignment. Here, we show that it is also possible to derive such equations for a more complex model of self-propelled particles with selective attraction-repulsion interaction and establish a direct link between the microscopic and macroscopic level of description. The model exhibits a large variety of different phases and we believe, it might be not only of interest from the biological point of view, as an alternative to models including explicit alignment of individual agents, but that it offers also interesting playground for the study of self-organization, pattern formation and phase transitions at far-from-equilibrium conditions.
 
\clearpage
	
%
\appendix
\section{Schematic Visualization of Social Interactions}
\label{app:visual}

 \begin{figure}[h]
  \begin{center}
   \includegraphics[width=0.32\linewidth]{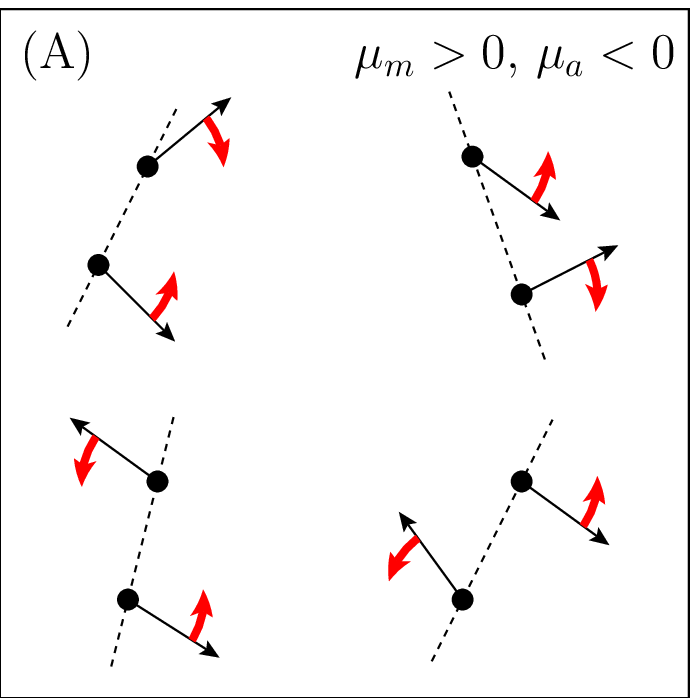}
   \includegraphics[width=0.32\linewidth]{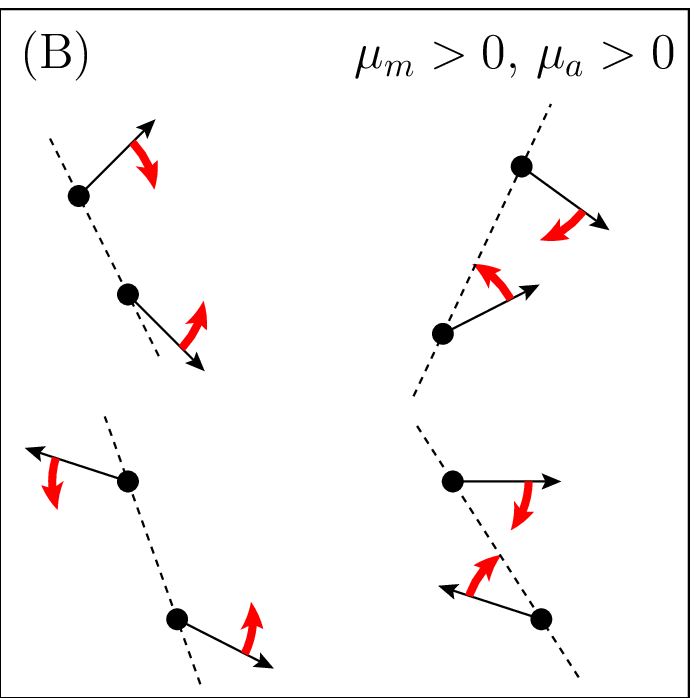} \\
   \includegraphics[width=0.32\linewidth]{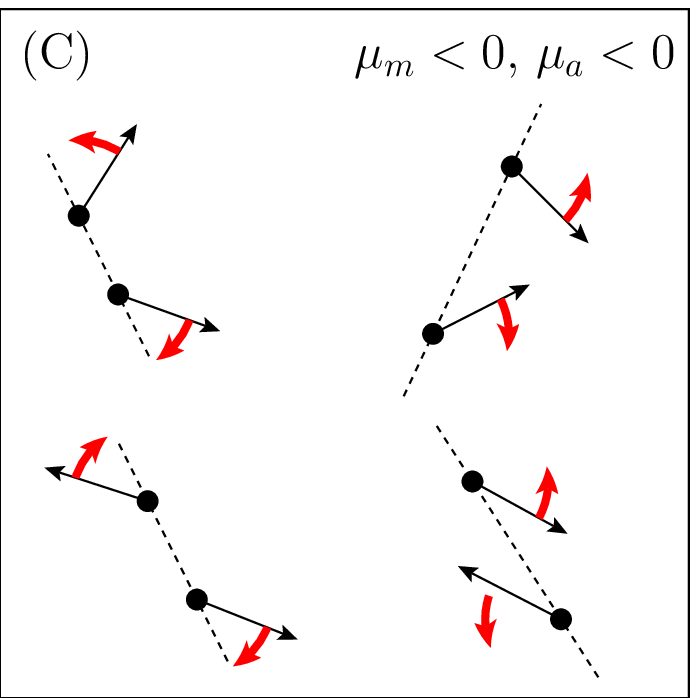}
   \includegraphics[width=0.32\linewidth]{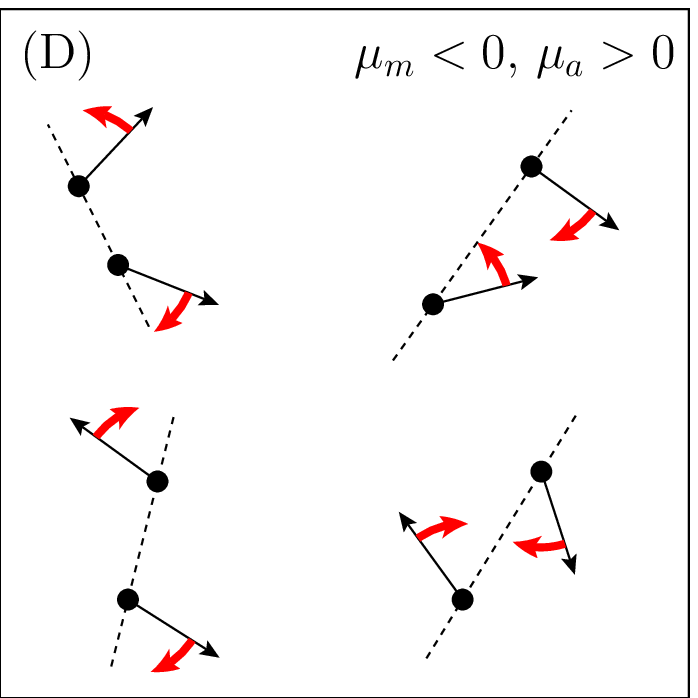}
   \caption{Turning of self-propelled particles due to the social interaction in the four regions of the  $(\mu_m,\mu_a)$-parameter space. In each panel, the left drawings corresponds to movement away $\tilde{v}_{ji}>0$ and right drawings to approach $\tilde{v}_{ji}<0$ of the agents. The red arrows indicate the torques acting on the agents, whereas the dashed line indicates the axis connecting the interaction partners, (anti-) parallel to the social forces; (A) effective alignment, (B) pure attraction, (C) pure repulsion and  (D) effective anti-alignment. 
   \label{fig:scheme}}
\end{center}
\end{figure}

\section{Calculation of $\tilde{K}_{p,n}$ \eqnref{eqn:fourierCoefficients:n:k}}
\label{sec:app:deriv:force:matrix}

In this section, it is sketched, how the matrix elements $\tilde{K}_{p,n}(\bi{k})$, defined by 
\Anumparts
 \begin{eqnarray}
   \fourier{K}{p}(\bi{k},t) &= \int \rmd^2 r \int_0^{2 \pi} \rmd \varphi \, F(\bi{r},\varphi,t) e^{i \bi{k}\cdot \bi{r} + i p \varphi} \, ,\\ 
   \fourier{K}{p}(\bi{k},t) &= \sum_{r=-\infty}^{\infty} \tilde{K}_{p,r}(\bi{k}) \fourier{g}{r}(\bi{k},t) \label{eqn:appendix:lin:comb:force:fourier}
 \end{eqnarray}
\endAnumparts
and used for the stability analysis of the spatial homogeneous solutions \eref{eqn:spatially:homogeneous:ordered:solution} of the Fokker-Planck equation \eref{eqn:fpe:particle:density} in section \ref{sec:stability:in:fourier}, are calculated. $\tilde{K}_{p,n}$ is obtained by inserting the inverse Fourier transform of the one-particle PDF
  \begin{eqnarray}
    p(\bi{r},\varphi,t) = \frac{1}{(2 \pi)^3} \sum_{n=-\infty}^{\infty} \int \rmd^2 k \, \fourier{g}{n}(\bi{k},t) e^{-i \bi{k}\cdot \bi{r} - i n \varphi}    \label{eqn:fourier:series:p}
  \end{eqnarray}
into \eqnref{eqn:self:consistent:force} and solving the remaining integral: 
 \begin{eqnarray}
				\fl F_{\varphi}(\bi{r},\varphi,t) =& \frac{1}{(2 \pi)^3} \sum_{p=-\infty}^{\infty} \int \mbox{d}^2 k \, \fourier{g}{p}(\bi{k},t) e^{-i \bi{k}\cdot \bi{r}} \int_{\varphi}^{\varphi + 2 \pi} \mbox{d}\varphi_j e^{- i p \varphi_j} \left \{ 2 s_0 \sin \left ( \frac{\varphi - \varphi_j}{2} \right)\right. \, \nonumber \\
    \fl & \int_{l_c}^{l_s} \mbox{d}r_{ji} \, r_{ji}  \left [ \mu_m(r_{ji}) \int_{\frac{\varphi + \varphi_j}{2}}^{\frac{\varphi + \varphi_j}{2} + \pi} \mbox{d}\alpha \, e^{-i \bi{k}\cdot \bi{r}_{ji}} \sin(\alpha - \varphi) \sin \left ( \frac{\varphi+ \varphi_j}{2} - \alpha \right) \right. \nonumber \\
				\fl & \left. - \mu_a(r_{ji}) \int_{\frac{\varphi + \varphi_j}{2}+\pi}^{\frac{\varphi + \varphi_j}{2} + 2 \pi} \mbox{d}\alpha \, e^{-i \bi{k}\cdot \bi{r}_{ji}} \sin(\alpha - \varphi) \sin \left ( \frac{\varphi+ \varphi_j}{2} - \alpha \right) \right] \nonumber \\
    \fl & \left. - \int_0^{l_c} \rmd r_{ji} \, r_{ji} \, \mu_r(r_{ji}) \int_0^{2 \pi} \rmd \alpha \, \sin (\alpha - \varphi) e^{-i \bi{k}\cdot \bi{r}_{ji}} \right \}  \, .
 \end{eqnarray}
In order to perform the integration over $\alpha$ and $\varphi_j$, the following identity is used to express the exponential function 
 \begin{eqnarray}
   e^{-i \bi{k}\cdot \bi{r}_{ji}} = J_0(k \, r_{ji}) + 2 \sum_{s=1}^{\infty} (-i)^s J_s(k \, r_{ji}) \cos \left (s \left ( \alpha-\chi \right) \right) \, ,
 \end{eqnarray}
where $\bi{k}=k\left (\cos \chi, \sin \chi \right)$ and $\bi{r}_{ji} = r_{ji} \left ( \cos \alpha, \sin \alpha \right)$. The Bessel functions of the first kind are denoted by $J_{\nu}(x)$.

Performing the integration yields the force expanded into a Fourier series and as a linear combination \eref{eqn:appendix:lin:comb:force:fourier} of the Fourier coefficients $\fourier{g}{r}(\bi{k},t)$, 
%
where one can read off the elements of the infinite dimensional matrix $\tilde{K}_{p,n}(\bi{k})$: 
 \begin{eqnarray}
   \label{eqn:matrix:of:forces}
   \fl \tilde{K}_{p,n}(\bi{k}) = & \frac{s_0 \pi^2}{2 i} \left \{ \left (\tilde{\mu}_{m,0}(k)-\tilde{\mu}_{a,0}(k) \right) \left ( \delta_{p,1} \delta_{n,1} - \delta_{p,-1} \delta_{n,-1} \right) + \left (\tilde{\mu}_{m,2}(k)-\tilde{\mu}_{a,2}(k) \right) \right. \nonumber \\
   \fl &\left. \left [ e^{-2i\chi} \left ( \delta_{p,-1} \, \delta_{n,1} - \delta_{p,-2} \, \delta_{n,0} \right) - e^{2 i \chi} \left ( \delta_{p,1} \, \delta_{n,-1} - \delta_{p,2} \, \delta_{n,0} \right) ...  \right ] \right \} \nonumber \\
   \fl & + 2 \pi^2 \tilde{\mu}_r(k) \, \delta_{n,0} \left ( e^{i \chi} \, \delta_{p,1} - e^{-i \chi} \, \delta_{p,-1} \right) + 2 s_0 \sum_{s=0}^{\infty} (-1)^s  \left ( \tilde{\mu}_{m,2s+1}(k) + \tilde{\mu}_{a,2s+1}(k) \right) \nonumber \\
   \fl & \left [ \Lambda_{p,s} \, e^{-i(2s+1)\chi} \, \delta_{p+2s+1,n} - \Gamma_{p,s} \, e^{i(2s+1)\chi} \, \delta_{p-(2s+1),n} \right ] \, .
 \end{eqnarray}
The following coefficients were introduced: 
\Anumparts
 \begin{eqnarray}
   \fl \tilde{\mu}_{m,\nu}(k)   &= \int_{l_c}^{l_s} \rmd r_{ji} \, r_{ji} \, \mu_{m}(r_{ji}) J_{\nu}(k \, r_{ji}) \, , \\
   \fl \tilde{\mu}_{a,\nu}(k)   &= \int_{l_c}^{l_s} \rmd r_{ji} \, r_{ji} \, \mu_a(r_{ji}) J_{\nu}(k \, r_{ji}) \, , \\
   \fl \tilde{\mu}_r(k) &= \int_0^{l_c} \rmd r_{ji} \, r_{ji} \, \mu_r(r_{ji}) J_1(k \, r_{ji})         \, , \\
   \fl \Lambda_{p,s} &= \frac{32 p}{(2s-1)(1+2s)(3+2s)(-1+2p+2s)(1+2p+2s)(3+2p+2s)}  \, , \\
   \fl \Gamma_{p,s}  &= \frac{32 p}{(2s-1)(1+2s)(3+2s)(-1+2p-2s)(1+2p-2s)(-3+2p-2s)} \, . 
 \end{eqnarray}
\endAnumparts
The Bessel functions $J_{\nu}(x)$ are oscillating functions with alternating Taylor coefficients. A truncation of the Taylor series (as done in section \ref{sec:hydrodynamic}) at a finite order may therefore lead to unphysical behaviour for large wavenumbers $k$. 


\section{Algebraic Sign of the Transport Coefficients}
\label{sec:App:transport:coefficients}
Besides the coefficient $\eta_1$, which is always positive, all transport coefficients may be of either sign, positive or negative. In the following, the sign of every coefficient is discussed in the $(\mu_m,\mu_a)$-parameter plane. For simplicity, it is assumed, that the interaction strengths $\mu_a$ and $\mu_m$ do not depend on the distance between two particles. 

\begin{figure}
\begin{center}
\centering\includegraphics[width=0.9\textwidth]{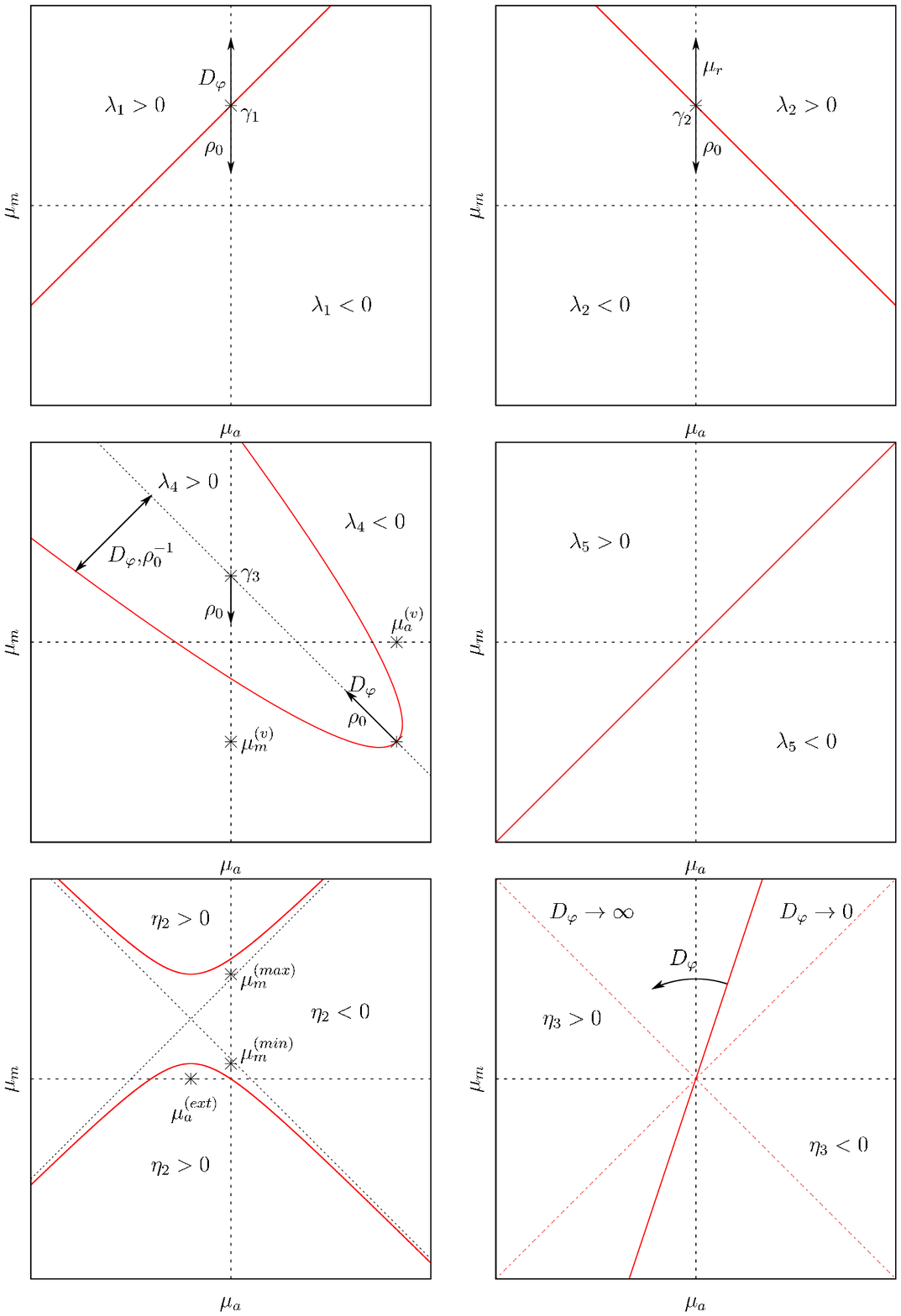}
\caption{The figure shows different regions in the $(\mu_m,\mu_a)$-parameter space, where the transport coefficients have different algebraic signs. Critical lines, where the transport coefficients vanish, are drawn in red. Black arrows indicate the shift of characteristic points with increasing noise $D_{\varphi}$ and density $\rho_0$, respectively. See the main text for definitions. }
\label{fig:VZ:kin:coefficients}
\end{center}
\end{figure}

$\lambda_1$ is positive above a critical line (positive slope and ordinate-intercept $\gamma_1$) which is given by
 \begin{eqnarray}
     \label{eqn:App:lambda:1:crit}
     \mu_m - \mu_a = \frac{8 D_{\varphi}}{\pi \rho_0 s_0^2 (l_s^2 - l_c^2)} \; {=:} \; \gamma_1 > 0 \, ,
 \end{eqnarray}
whereas $\lambda_2$ is positive above a critical line (negative slope and positive ordinate-intercept $\gamma_2$), given by
 \begin{eqnarray}
     \fl \mu_m + \mu_a = \frac{27 \pi}{16 s_0 \rho_0 (l_s^3 - l_c^3)} \left ( \frac{s_0^2}{2} + \frac{\pi \rho_0 \mu_r l_c^3}{6} \right) \; {=:} \; \gamma_2 \ge \frac{9 \pi}{32 s_0} \frac{l_c^3}{l_s^3 - l_c^3} \, \mu_r >  0 \, .
 \end{eqnarray}
The analysis of the coefficient $\lambda_4$ is somewhat more involved. $\lambda_4$ is greater than zero inside a parabola, which is oriented along the line
 \begin{eqnarray}
     \label{eqn:App:lambda:3:crit}
     \mu_m + \mu_a = \frac{405 \pi s_0}{256 \rho_0 \left ( l_s^3 - l_c^3 \right)} \; {=:} \; \gamma_3 >  0 \, .
 \end{eqnarray}
The width of the parabola is proportional to the noise strength $D_{\varphi}$ and inverse proportional to the density $\rho_0$. The vertex of the parabola has the coordinates
\Anumparts
 \begin{eqnarray}
     \mu_a^{(v)} = \frac{1}{2} \left ( \frac{405 \pi s_0}{256 \rho_0 (l_s^3 - l_c^3)} + \frac{25 s_0^4}{2 \pi \rho_0 (l_s^4-l_c^4)  D_{\varphi}} \right )    \, , \\
     \mu_m^{(v)} = \frac{1}{2} \left ( \frac{405 \pi s_0}{256 \rho_0 (l_s^3 - l_c^3)} - \frac{25 s_0^4}{2 \pi \rho_0 (l_s^4-l_c^4) D_{\varphi}} \right )    \, .
 \end{eqnarray}
\endAnumparts
The parabola itself is given by\footnote[7]{Equations \eref{eqn:App:parabola:lambda_4} and \eref{eqn:App:hyperbola:eta2} are much easier to visualize in new coordinates $\beta_1=\mu_m+\mu_a$ and $\beta_2 = \mu_m - \mu_a$. The transformation basically corresponds to a rotation in the $(\mu_m,\mu_a)$-parameter plane. }
 \begin{eqnarray}
   \label{eqn:App:parabola:lambda_4}
   \fl   \mu_m - \mu_a = \frac{8 s_0^4}{\pi \rho_0 (l_s^4 - l_c^4)  D_{\varphi}} &\left [ \left ( \frac{64 \rho_0 (l_s^3 - l_c^3) }{135 \pi s_0} \right)^2 (\mu_m + \mu_a)^2 \right. \nonumber \\ 
   \fl & \left. \; - \frac{32 \rho_0 (l_s^3 - l_c^3) }{45 \pi s_0} \, (\mu_m +  \mu_a)  - 1    \right ]    \, .
 \end{eqnarray}
The coefficient $\lambda_5$ is positive above the main diagonal, i.e. $\mu_m>\mu_a$. Transport coefficient $\eta_2$ is less than zero inbetween two hyperbolas, which are bounded by the asymptotes \eref{eqn:App:lambda:1:crit} and \eref{eqn:App:lambda:3:crit}. The hyperbolas are determined by
 \begin{eqnarray}
   \label{eqn:App:hyperbola:eta2}
     \mu_m-\mu_a = \gamma_3 \, \frac{\left (\mu_m + \mu_a\right)}{\left (\mu_m + \mu_a\right) - \gamma_1}
 \end{eqnarray}
or rather
 \begin{eqnarray}
     \mu_m = \frac{1}{2} \left (\gamma_1 + \gamma_3 \pm \sqrt{4 \mu_a^2 + 4 \mu_a (\gamma_1 - \gamma_3) + (\gamma_1 + \gamma_3)^2} \right)   \, .
 \end{eqnarray}
The extrema of the hyperbolas are located at
\Anumparts
 \begin{eqnarray}
     \left (\mu_a^{(ext)},\mu_m^{(min)} \right ) = \left ( \frac{\gamma_3 - \gamma_1}{2} , \frac{\left ( \sqrt{\gamma_1} - \sqrt{\gamma_3} \right)^2}{2} \right)    \, , \\
     \left (\mu_a^{(ext)},\mu_m^{(max)} \right ) = \left ( \frac{\gamma_3 - \gamma_1}{2} , \frac{\left ( \sqrt{\gamma_1} + \sqrt{\gamma_3} \right)^2}{2} \right)    \, . 
 \end{eqnarray}
\endAnumparts
The critical line for $\eta_3 = 0$ depends on the noise strength, whereby the critical noise strength reads
 \begin{eqnarray}
   D_{\varphi}^{(c)} = \frac{675 \pi^2 s_0^3 (l_s^2-l_c^2)}{2048 (l_s^3-l_c^3)} \, .  
 \end{eqnarray}
$\eta_3$ is positive, if
 \begin{eqnarray}
   \mu_m \cases{ > \mu_a \cdot  \frac{D_{\varphi}^{(c)} + D_{\varphi}}{D_{\varphi}^{(c)} - D_{\varphi}} \, , \quad D_{\varphi} < D_{\varphi}^{(c)} \, , \\  < \mu_a \cdot \frac{D_{\varphi}^{(c)} + D_{\varphi}}{D_{\varphi}^{(c)} - D_{\varphi}} \, , \quad D_{\varphi} > D_{\varphi}^{(c)} \, . } 
 \end{eqnarray}
Please note, that the slope of the lines is bounded by one ($D_{\varphi} = 0$) and minus one ($D_{\varphi} \rightarrow \infty$), respectively. Hence, $\eta_3$ is always positive for $\left \{ (\mu_a,\mu_m): \mu_a<0 \, , -\abs{\mu_a} < \mu_m < \abs{\mu_a} \right \}$ and $\eta_3$ always negative for $\left \{ (\mu_a,\mu_m): \mu_a>0 \, , -\mu_a < \mu_m < \mu_a \right \}$.  All criteria discussed above are summarized in figure \ref{fig:VZ:kin:coefficients}.  

The stability of the hydrodynamic equations \eref{eqn:hydrodynamic:theory:direct} requires that both $\lambda_4$ and $\lambda_5$ are greater than zero. This is in principle true in the effective-alignment region (at least for sufficient high noise and low densities), as well as parts of the pure attraction and pure repulsion region. In this parameter region, the hydrodynamic equations describe collectively moving bands. 

\section{Temperature Tensor}
\label{sec:A:Temperature}

In \cite{romanczuk_mean-field_2012,grossmann_active_2012,romanczuk_swarming_2012}, the symmetric temperature tensor 
\begin{eqnarray}
 \fl \mbox{
			$\hat{T} = \left ( 
				 \begin{array}{cc}
 					\mean{\cos \varphi^2} - \mean{\cos \varphi}^2 & \mean{\sin \varphi \cos \varphi} - \mean{\sin \varphi}\mean{\cos \varphi}\\ 
 					\mean{\sin \varphi \cos \varphi} - \mean{\sin \varphi}\mean{\cos \varphi} & \mean{\sin \varphi^2} - \mean{\sin \varphi}^2 
				\end{array}
			 \right ) 
			$
		}
\end{eqnarray}
is defined. It is a measure for the width of the velocity distribution (fluctuations around the mean velocity). The temperature is basically related to the second Fourier coefficients:
\Anumparts
\begin{eqnarray}
  T_{xx} &= \frac{1}{2}\left (1 + \frac{\fourier{f}{2}(\bi{r},t)+\fourier{f}{-2}(\bi{r},t)}{2 \rho(\bi{r},t)} \right) - \left ( \frac{\fourier{f}{1}(\bi{r},t) + \fourier{f}{-1}(\bi{r},t)}{2 \rho(\bi{r},t)} \right)^2 \, , \\
  T_{xy} &= \frac{\fourier{f}{2}(\bi{r},t)-\fourier{f}{-2}(\bi{r},t)}{4 i \rho(\bi{r},t)} - \frac{\left (\fourier{f}{1}(\bi{r},t)\right)^2 - \left (\fourier{f}{-1}(\bi{r},t)\right)^2}{4 i \left (\rho(\bi{r},t)\right)^2} \, , \\
  T_{yy} &= \frac{1}{2}\left (1 - \frac{\fourier{f}{2}(\bi{r},t)+\fourier{f}{-2}(\bi{r},t)}{2 \rho(\bi{r},t)} \right) + \left ( \frac{\fourier{f}{1}(\bi{r},t) - \fourier{f}{-1}(\bi{r},t)}{2 \rho(\bi{r},t)} \right)^2 \, .
\end{eqnarray}
\endAnumparts
The second Fourier coefficients are assumed to be fast variables in section \ref{subsec:hydrodynamic_limit}, so that the temperature is implicitly contained in the hydrodynamic description of the system, even though the dynamics of the temperature tensor is not derived explicitly in this context.

\section*{References} 

\bibliographystyle{unsrt}	

\end{document}